\pdfoutput=1
\documentclass[smallextended]{svjour3}       % onecolumn (second format)
\smartqed  % flush right qed marks, e.g. at end of proof
\usepackage{graphicx}
\journalname{Space Science Reviews}
\usepackage{natbib}
\usepackage{epsfig}
\usepackage{color}
\usepackage{times}
\usepackage{amssymb,amsmath}
\usepackage{multirow}
\usepackage{boldline}
\usepackage{enumerate}
\begin{document}

\title{The origin, early evolution and predictability of solar eruptions}

\titlerunning{The origin of solar eruptions}    

\author{Lucie M. Green \and
        Tibor T\"{o}r\"{o}k \and 
	Bojan Vr\v{s}nak \and
	Ward Manchester IV \and
	Astrid Veronig
}

\institute{Lucie M. Green \at
              UCL, Mullard Space Science Laboratory, Holmbury St. Mary, RH5 6NT, UK \\
              \email{lucie.green@ucl.ac.uk}           %  \\
           \and
           Tibor T\"{o}r\"{o}k \at
              Predictive Science Inc. 9990 Mesa Rim Rd, Ste 170, San Diego, CA 92121, USA \\
	      \email{tibor@predsci.com}           %  \\
           \and
	   Bojan Vr\v{s}nak \at
		Hvar Observatory, Faculty of Geodesy, University of Zagreb, Kaciceva 26, 10000 Zagreb, Croatia \\
	        \email{bvrsnak@gmail.com}             %  \\
           \and
	   Ward Manchester \at
 		Department of Climate and Space Sciences and Engineering, University of Michigan, Ann Arbor, MI 48109, USA \\
	        \email{chipm@umich.edu}             %  \\
           \and
	   Astrid Veronig  \at
  		Institute of Physics/IGAM, University of Graz, Universitaetsplatz 5, A-8010 Graz, Austria \\
	        \email{astrid.veronig@uni-graz.at} 
}

\date{Received: date / Accepted: date}

\maketitle

\begin{abstract}

Coronal mass ejections (CMEs) were discovered in the early 1970s when space-borne coronagraphs revealed that
eruptions of plasma are ejected from the Sun. Today, it is known that the Sun produces
eruptive flares, filament eruptions, coronal mass ejections and failed eruptions; all thought
to be due to a release of energy stored in the coronal magnetic field during its drastic reconfiguration. 
This review discusses the observations and physical mechanisms behind this eruptive activity, 
with a view to making an assessment of the current capability of forecasting these events for space
weather risk and impact mitigation. Whilst a wealth of observations exist, and detailed models have been
developed, there still exists a need to draw these approaches together. In particular more realistic 
models are encouraged in order to asses the full range of complexity of the solar atmosphere and the criteria
for which an eruption is formed. From 
the observational side, a more detailed understanding of the role of photospheric flows and reconnection is needed
in order to identify the evolutionary path that ultimately means a magnetic structure will erupt.

\end{abstract}

\section{Introduction}
\label{intro}

Our Sun is a dynamic star, exhibiting a range of large-scale eruptive activity that, over the last roughly 150 years of study,
has led to various activity categories being developed. These include eruptive flares, 
filament eruptions, coronal mass ejections (CMEs)
and failed eruptions. The modern understanding is that this list represents 
different manifestations of a disruption and reconfiguration of
the coronal magnetic field, which ultimately leads to an ejection of magnetised plasma that may, 
or may not, propagate into the heliosphere.
In this review we use the general term CME to encapsulate all eruptive activity types, and focus on the 
physical processes that are involved.

\begin{figure*}[ht!]
\begin{center}
\includegraphics[width=1.\textwidth]{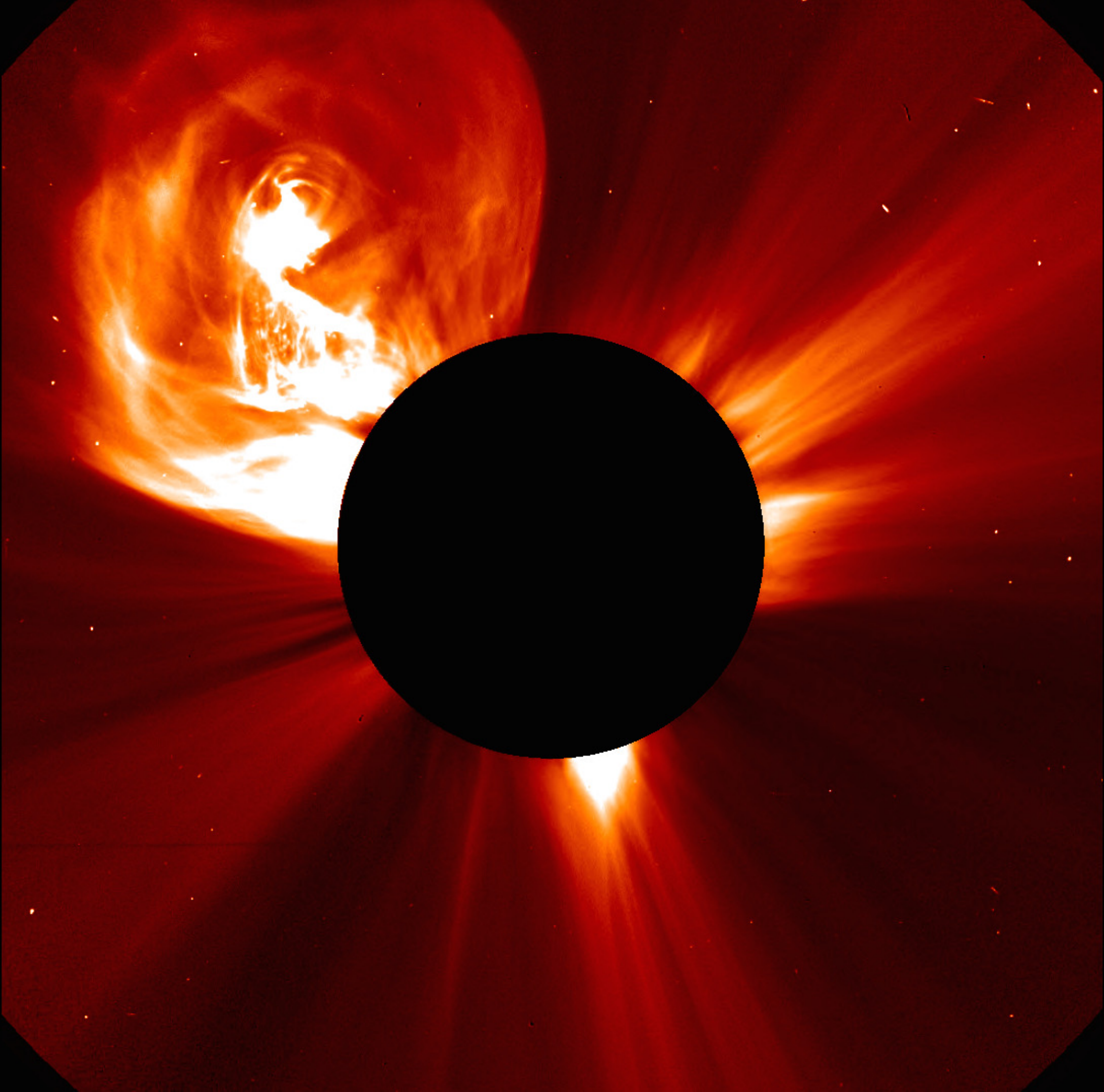}
\end{center}
\caption{
A coronal mass ejection imaged by the SOHO/LASCO 
C2 coronagraph on 4th January 2002. The CME is revealed due to Thomson scattered photospheric light 
and the ejection can be seen in the upper left of the image to have a bright core in addition to
the surmounting structure. The black disk is a result of the occulting disk, which 
has a radius of $1.7 R_{sun}$.
}
\label{cme_lasco}
\end{figure*}

Coronal mass ejections are the sudden release of $10^{15}$--$10^{16}$g of chromospheric
and coronal plasma into the heliosphere. Their speed ranges from between a few tens to a few
thousands km\,s$^{-1}$, with an average value of $\sim490$\,km\,s$^{-1}$
\citep{webb12}. Fast CMEs (that have speeds greater than the background solar wind
through which they are propagating) can drive shocks ahead of them.
CMEs are identified through the use of coronagraph data in which they are
seen via the detection of Thomson scattered photons in outward moving plasma
structures that have a higher electron density than the background corona (Figure \ref{cme_lasco}). Their plane-of-sky
configuration varies from event to event and includes amorphous blobs, a ``3-part'' structure
consisting of a bright front, dark cavity and bright central
core \citep{illinghundhausen85}, and helical shapes. The plasma traces out the structure of the
magnetic field and the ejected plasma is carried away from the Sun as it is ``frozen-in'' to
the magnetic field. 
Indeed studying the magnetic field is central to understanding CMEs since
the energy to power CMEs, which can be up to the
order of $10^{32}$ ergs, 
can only come from the conversion of free magnetic energy into other forms
\citep{forbes2000}.
The free magnetic energy is the excess energy stored in the coronal field above the energy of the 
current-free potential field configuration. The free magnetic energy is
that available to be released and is stored in the form of
electric currents in the pre-eruptive non-potential magnetic structures 
\citep[for a review see, e.g.,][]{low96, priest02,vrs06asr,aulanier14,schmieder15}.

%%------------------------------ FIGURE --------------------------------
%%-----------------------------------------------------------------------

When CMEs are formed at the solar limb many can be seen to originate from
coronal streamers \citep{hundhausen1993}. However, for the majority of CMEs, the eruption includes a
low-lying sheared core field, which may or may not contain filament plasma, and that 
cannot be observed with a coronagraph. Therefore
instruments that image the solar disk must be used to study the origin of CMEs in addition
to coronagraph data. It is these low altitude observations that showed CMEs are
closely related to solar flares \citep[e.g.][]{vrs08angeo,schmieder15}. 
During a solar flare, free magnetic energy is converted into energetic particles, heat and
electromagnetic radiation and the close temporal and spatial CME-flare relationship shows that they are the 
product of the same disruption and reconfiguration of
the coronal magnetic field \citep[e.g][and references therein]{temmer10}. 
A relationship between
CME occurrence and flare size has been noted, with observations revealing that the
likelihood of a flare having an associated CME increases with flare magnitude and that all
flares above X5 in the GOES classification occur in concert with a CME 
\citep{andrews03,yashiro05}.
However, it should be pointed out that CMEs and flares do not occur on a one-to-one basis.

CMEs are common events and their occurrence rate varies with the phase of the solar cycle.
On average one CME is formed per day at cycle minimum with around five per day at cycle
maximum \citep{webb12} when active regions are more abundant on the solar disk.
Active regions can be the source of CMEs during their emergence phase, when
current carrying magnetic field is being brought into the corona \citep{wang94,leka96}, and through
the decay phase when fragmentation and flux dispersal occur, redistributing the magnetic field
into the quiet sun. However, CME characteristics can be very different at different evolutionary stages
of an active region. From compact, fast CMEs that occur during the emergence phase to quiescent filament
eruptions that occur when active regions are at the end of their lives, and finally streamer blowouts associated 
with the largest-scale coronal structures/or quiescent filaments. The eruptive structure that forms the CME can come from within 
an active region and also between active regions. 
Two examples of how the CME \textit{rate} varies with an active region's evolutionary phase were presented in
\cite{demoulin02} and \cite{green02}. These studies followed NOAA 7978 and NOAA AR 8100, respectively,
for several months and showed that these active regions were CME productive throughout their lifetime
(whereas flare activity disappeared as the decay phase progressed).
The findings indicate that CMEs do not seem to be so dependent on the
magnetic field conditions that are necessary for flares to occur. Flares, especially major flares, 
occurred when each active region was young and flux emergence had recently taken place. 
CMEs on the other hand occurred at all evolutionary stages, including during
the decay phase when there was no new major flux emergence and the energy density of the 
magnetic field had reduced.
These studies revealed that CMEs are not as dependent on strong field and magnetic complexity as are flares.
Still, active regions that have large and complex magnetic field configurations can be the
origin of many CMEs during their emergence phase. For example, NOAA active region 9236
produced 17 CMEs in 5 days \citep{gopalswamy05} and NOAA active region 8100 produced
15 CMEs in 7 days \citep{green02}. However, these large and dramatic regions should
not distract from the fact that CMEs are formed in magnetic structures that span a range of spatial scales.
The size-scale of the
pre-eruption structure can be as large as 100s thousands of km long (polar crown filaments)
but the structure can be as small as 
only 10s of thousands of km across
\citep{mandrini05}.

Some CMEs detected in coronagraph data show no signatures
in the lower atmosphere, earning them the name “stealth CMEs”. They originate at altitudes
of $>$ 0.1 $R_\odot$, where $R_\odot$ is the solar radius, \citep{robbrecht09} 
in magnetic structures where the field
strength is weak and the plasma density low, meaning that observational signatures are
less likely to be seen \citep{howard13}. A study of the Sun at solar minimum
suggests that stealth CMEs may make up 30\% of all CMEs that occur \citep{ma10}.
\citet{kilpua14} found in their study after solar maximum in 2014 that ten out of the 16 studied events could
be classed as stealth CMEs. 

Taking into account both the remote sensing and in situ data it has been realised that a CME 
in the inner corona can be
composed of some or all of the following:

\begin{itemize}
\item a shock that moves ahead of the CME
\item a frontal structure with high plasma density ($n_e \approx 10^{14}$ m$^{-3}$),
coronal temperature ($\sim$2\,MK) and high magnetic field strength ($10^{-4}$ T)
\item a cavity with low plasma density ($n_e \approx 10^{13}$ m$^{-3}$), coronal temperature (~2\,MK)
and field strength of few $10^{-4}$ T
\item a prominence core with high plasma density ($n_e \approx 10^{17}$ m$^{-3}$),
low temperature ($\sim$80000\,K) and high field strength (few $10^{-3}$ T)
\item a post-eruption arcade with a plasma temperature
of $\sim$10\,MK and a few $10^{-4}$ T field strength
\end{itemize}

Studies of CMEs have found a new relevance in the last three decades through the realisation
that they can drive severe space weather at Earth \citep{gosling93,green15}.
Parameters that are important for determining the level to which a CME will be geo-effective
upon its impact with the magnetosphere include the strength of the southward component of the CME's
magnetic field ($B_z$), the speed and plasma density of the CME ($V_{\mathrm CME}$, $\rho$) and the
CME's dynamic pressure ($\rho\,V_{CME}^{2}$). 
This means that fast CMEs with a strong
and sustained southward magnetic field component and/or a high plasma density
are likely to be the ones that have the most significant effect. Historical case studies of such
events include the Carrington event of 1859 when a CME reached the Earth in 
just 17.5 hours \citep{cliver04}
giving it an average velocity along the Sun-Earth line of $\approx 2300\,\mathrm{km\,s}^{-1}$, 
and the March 1989 event which was
exceptionally bright in coronagraph data, implying it had a high plasma density although
it also had a much lower average Sun--Earth line speed  of around 770\,km\,s$^{-1}$ 
\citep{feynman94}.
These CMEs produced geomagnetic storms with a disturbance storm time ($Dst$) index 
estimated to be
-850 nT \citep{siscoe2006b} and -548 nT \citep{cliver04} respectively. It is worth noting that
the $Dst$ value during the 1859 event has been a topic of controversy, 
with \cite{tsurutani2003} proposing that -1760 nT may have been reached. 

Especially important in the creation of severe geomagnetic storms appear to be CMEs that
follow one another in quick succession \citep{liu2014,dumbovic2015,temmer2015,vennerstrom2016}. 
The Carrington event is one example of such a case.
In these events, the preceding CME may ``pre-condition'' the space between the Sun and the Earth
so that the following CME is embedded in a faster solar wind stream and so experiences
less drag; allowing it to maintain a high velocity out to 1AU \citep{temmer2017}.

The impact of a CME extends beyond its direct interaction with the magnetosphere as
shocks driven by fast CMEs \cite[roughly above around 800\,km\,s$^{-1}$,][]{kahler2001} can
accelerate particles, producing an additional space weather effect
when these solar energetic particles (SEPs) arrive at the Earth \citep{reames2013}.
Meaning that when CME predictions are being developed, speed and spatial extent of the CME
are important quantities to predict too. As are the upstream wind conditions
ahead of the shock that can affect the shock geometry. 

The very significant role that CMEs play in the creation of space weather has naturally led to a
desire to predict their time of arrival at Earth. 
Once a CME has been launched, it can take from 
less than a day to more than three days before it arrives. Since the fastest CMEs can
be the most geo-effective, the ability to predict their occurrence before they happen would
significantly improve forecasting lead times. 
In this review paper we draw together the wide-ranging observational (Section \ref{sec:origin_obs})
and theoretical (Section \ref{sec:origin_theo_sim}) work 
on CMEs to assess the current status of understanding how eruptive structures are formed and
how energy is injected into the magnetic field, over timescales
of weeks to hours before the eruption.
We focus on processes that occur during the energy build-up phase that bring the magnetic field
to a point where a "driving" mechanism occurs to drive the rapid expansion of the structure.
From this, we look at current and future directions of predicting CMEs based 
on both observations and modelling (Section \ref{sec:predict}).

\section{Origin and evolution of solar eruptions}
\label{sec:origin}

\subsection{Observations}
\label{sec:origin_obs}

\begin{figure*}[hb!]
\begin{center}
\includegraphics[width=1.\textwidth]{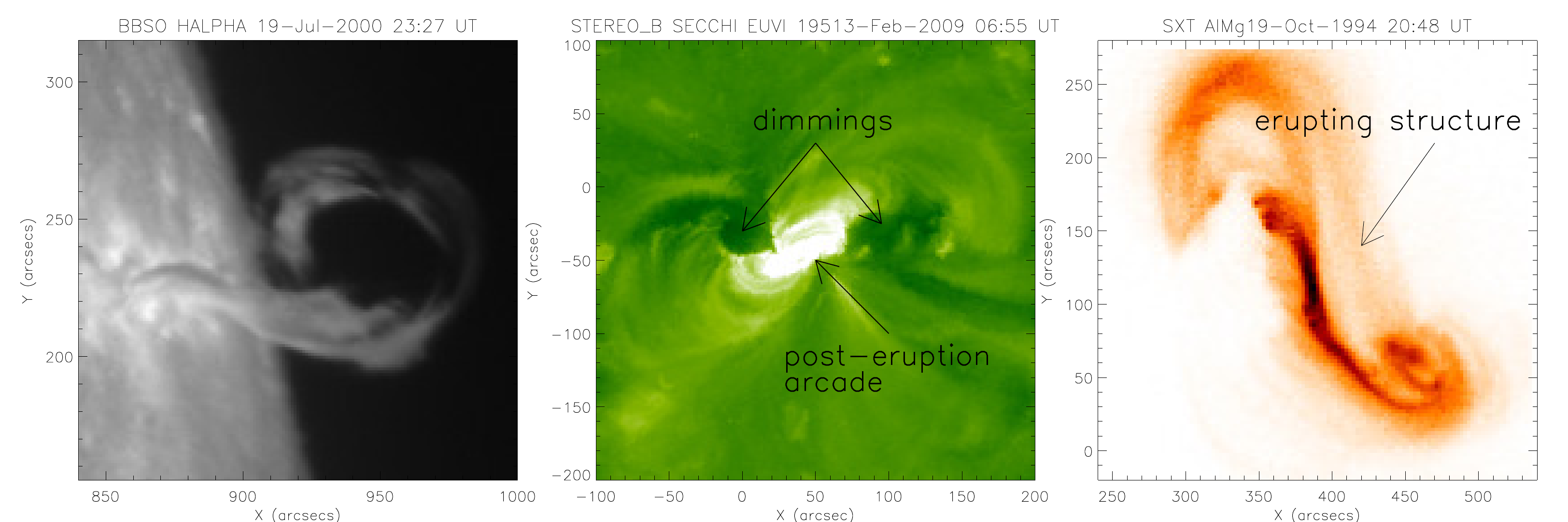}
\end{center}
\caption{
Lower coronal signatures of the occurrence of a coronal mass ejection (CME). There are a variety of observational
manifestations of CMEs including filament eruptions (left panel), post-eruption arcades (also called flare 
arcades, middle panel), dimming
regions (middle panel) and rising EUV/soft X-ray structures (right panel).
} 
\label{eruption_examples}
\end{figure*}

Coronal mass ejections are observed using a wide range of imaging and spectroscopic
data, spanning wavelengths from radio to X-ray and altitudes from the photosphere 
to the outer corona. 
These observations reveal that in many cases, the onset of a CME will be marked by the 
eruption of a filament/prominence in the lower solar atmosphere. 
Indeed, over 70\% of CMEs begin this way \citep{Munro79}.
The filament may be seen as a bright core in coronagraph data in CMEs that have a 
3-part structure. Lower coronal observations of CME source regions can also show the rise of 
plasma structures that are emitting at EUV and/or
soft X-ray wavelengths. Such rising structures have been seen in {\em Yohkoh}/SXT data 
\citep[e.g.][]{moore01,green14}, {\em SDO}/AIA 
channels \cite[e.g.][]{liu2010,zharkov11} and {\em STEREO}/EUVI data
\citep{bein11,patsourakos2013}.
In the wake of a CME, the reconfiguration results in a post-eruption 
arcade, which is also known as a flare arcade.
When the CME source region is located at the solar limb, hot plasma structures referred to as
plasmoids or flux ropes have been observed \citep{shibata1995,cheng2011,reeves11}.
Once the eruption is well underway, the footpoints
of the expanding and erupting magnetic volume can sometimes be identified through EUV or 
soft X-ray regions which have dimmed due to the reduction in plasma density 
\citep{rust76,sterling1997,zarro1999}. Figure \ref{eruption_examples} provides 
examples of the variety of lower corona
signatures that indicate the occurrence of a CME.

In this section we review observational aspects of the coupled evolution of the 
photosphere and the corona in the long-term (Section \ref{subsec:longterm_obs})
and short-term (Section \ref{subsec:shortterm_obs})
leading up to the eruption 
and during the rapid acceleration phase of the eruption itself (Section \ref{subsec:evo_eruption}).

%%------------------------------ FIGURE --------------------------------
%%-----------------------------------------------------------------------

\subsubsection{Long-term evolution prior to eruption}
\label{subsec:longterm_obs}

The energy required to power a CME appears to be built-up in the coronal field
over days/weeks prior to the eruption, in an energy storage and release process. 
This injection
of energy is driven by both the emergence of flux \citep[which is non-potential,][]{leka96} and 
photospheric motions due to convection and differential rotation that can stress the coronal field. 
In some cases sunspot rotation is 
observed \citep[e.g.,][]{brown2003} and/or rotational
motions of same-polarity magnetic fragments, one about the other \citep{james2017}.

Development of non-linear force-free (NLFFF) modelling techniques have provided methods to
reconstruct the coronal magnetic field using photospheric vector magnetograms. In turn
the evolution of the energy in the coronal field can be analysed from these extrapolated
configurations. The analysis of the magnetically complex
NOAA active region 11158 by \cite{sun2012} followed the 
energy stored in the coronal field over a period of five days, during which time the
region produced multiple flares and CMEs. The study found a rapid increase in 
the free magnetic energy of the coronal field lasting around four hours
as the flux emergence began.
This was followed by a more gradual injection of free energy into the coronal
configuration during the following $\sim$two days. The energy was mostly accumulated at low heights 
(less than 6 Mm) along a polarity inversion line where a low-lying filament formed,
and where later eruptions originated. 
The extrapolated coronal field configuration shows that around the time of an
X-class flare approximately 3.4$\times10^{31}$ ergs of energy is released. 
The NLFFF modelling studies by \cite{gibb14, mackay11} followed the evolution of two 
short-lived small bipolar active regions
during their emergence and decay phases using a magnetofrictional method. 
The simulated coronal field in these studies captured the energy build-up prior 
to the CMEs that the regions produce.
Free energy is found to be built-up during the flux emergence phase and also during the decay 
phase when the regions start to undergo flux cancellation at their internal polarity inversion 
lines (flux cancellation is described and discussed at the end of this section). 

As well as energetics, the quantity of magnetic helicity has been considered in the evolution
of active regions that produce CMEs. Magnetic helicity is the quantity that encapsulates
how twisted, sheared and interlinked the magnetic field is. 
Magnetic helicity is approximately conserved even during non-ideal magnetohydrodynamic (MHD)
processes \citep{berger84}. For example, magnetic reconnection serves to transport magnetic 
helicity to larger spatial scales, opposite to energy, which cascades to 
smaller scales where it can dissipate.
Magnetic helicity is injected into the corona via flux emergence, photospheric motions
and torsional and shear Alfv\'en waves 
and this, along with its conservation property, has led to the suggestion that CMEs may act as a 
valve to prevent magnetic helicity from endlessly accumulating in the solar atmosphere 
\citep{heyvaerts1984, low96}. Indeed, active regions during their entire lifetime
can be the source of many CMEs, which eject a significant amount of magnetic helicity 
\citep{demoulin02,green02}.
It has even been proposed that CMEs are the unavoidable product of magnetic helicity
accumulation in the force-free magnetic field of the corona \citep{zhang2006}.
The proposal of an intimate link between this topological quantity and a CME spurred a 
wave of research in this area. New methods to quantify magnetic helicity injection into the solar 
atmosphere via flux emergence and photospheric flows provided the tools to be able to make 
these investigations \citep{demoulin2003}.

Many studies of the evolution of magnetic helicity in active regions in the time periods around 
the occurrence of flares and CMEs have been carried out. \cite{nindos2004} studied 
the coronal magnetic helicity evolution of active regions that were the source of 133 M-class 
and X-class flares. They found that the magnetic helicity contained in the coronal field
of an active region was larger in regions that produced flares that had associated CMEs than for 
those regions that had CME-less flares. Other studies have looked at magnetic helicity flux, even proposing 
that a threshold helicity injection rate must be exceeded for a CME to occur \citep{labonte2007}.
However, in contrast to this, other work has found that major changes in the magnetic helicity 
content of an active region can occur both before and
after a CME \citep{smyrli2010}. Helicity injection prior to a CME in the \cite{smyrli2010}
study was seen to be due to flux emergence, whereas the changes after the CME were proposed to be due
to the torque imbalance between the helicity-depleted corona and the sub-photospheric 
portions of the flux as suggested in \cite{longcope2000,demoulin02,green02}.
Whether 
or not magnetic helicity is a quantity that can be used to predict CME occurrence will
be discussed further in Section \ref{sec:pedict_obs}. 

The role of flux emergence and photospheric shear flows in injecting sufficient helicity into the
corona before a CME
were studied by \cite{nindoszhang02}
using NOAA active region 9165. The study finds that the horizontal flows did not 
inject sufficient magnetic helicity
into the corona to account for that ejected by CMEs. Instead, flux emergence is invoked as the dominant
helicity source. 
However, studies of smaller and shorter-lived bipolar active regions
suggest that during the decay phase of these short-lived regions the photospheric motions caused by convection
(super-granular flows) may play an important role in injecting helicity into the corona
\citep{savcheva12,gibb14}. All the above studies indicate that the magnetic 
helicity in the corona can be injected via variety of mechanisms, and the mechanism that dominates
may depend on the evolutionary stage of the region.
Next, we focus on
how a ubiquitous process observed in the photosphere known as flux cancellation 
might prime the corona, ready for an eruption.

Flux cancellation is a common phenomenon that occurs all over the Sun \citep{livi1985}
and is 
the name given to an observational event consisting of the approach of two opposite 
polarity fragments that collide and then disappear from the photospheric 
magnetic field measurement \citep{martin1985}. 
It’s importance for this review paper
is that flux cancellation is often observed in the time leading up to a CME
\citep{martin1985}, which suggests that it may be important
in the creation of an eruptive magnetic field configuration and its destabilisation. 
The physical interpretation 
of flux cancellation is that when opposite polarities collide they undergo reconnection in the 
photosphere or chromosphere \citep[e.g.][]{yurchyshyn2001,litvinenko1999}. 
These opposite polarities represent the footpoints of two sheared loops and the 
low-altitude reconnection reconfigures the magnetic field so that a long loop
forms in the corona and also a very short loop that submerges below the photosphere
due to its tension force overcoming buoyancy.
Shear and free magnetic energy are concentrated along polarity inversion lines
where flux cancellation is ongoing for extended periods of time \citep{welsch2006}, 
hence it relevance for CME studies.

In some small bipolar active regions, an eruptive configuration is formed via flux cancellation
along the internal polarity inversion line 
during the decay phase of the region, when the flux is fragmenting.
This is well observed in bipolar active regions that form S-shaped “sigmoidal” emission structures in
soft X-ray or EUV wavebands over a couple of days \citep[e.g.][]{green11}.
Both the coronal evolution and the photospheric evolution in these cases match well the scenario proposed by
\cite{ballegooijen1989} in which a sheared arcade is transformed into a flux rope. 
This is also in line with the theoretical findings that
sigmoids are created by hot plasma along field lines in the vicinity of a quasi-separatrix layer
formed in flux rope topology \citep{titov99}.

In summary, the evolution of an active region prior to a CME involves the emergence of magnetic flux into 
the corona that is then modified by photospheric flows, which eventually disperse the field over an ever wider area.
During the lifetime of the active region flux cancellation may be occurring that
is able to modify the configuration of the magnetic field and allow free magnetic energy
to buildup along polarity inversion lines.
CMEs can occur at any point during this evolutionary sequence although the frequency of 
CMEs is highest at times of new flux emergence \citep{green03}.
 
\subsubsection{Short-term evolution prior to eruption}
\label{subsec:shortterm_obs}

Collated observations of the evolution of CME source regions in the tens of minutes before an eruption reveal
that the magnetic structure evolves through distinct dynamic phases. 
The majority of CMEs exhibit a slow-rise phase, a phase of rapid acceleration and then
a phase where the erupting structure propagates into the heliosphere 
\citep{zhang01,vrsnak01}. 
The slow rise phase occurs over
timescales of minutes to hours and can be observationally identified through the upward motion of a filament
or the expansion of coronal loops. The phase of rapid acceleration then occurs 
and signals that the eruption is in progress. During the propagation phase 
the CME interacts with the ambient solar
wind \citep{vrs08dij}. Here we describe the short-term evolution in the hours or
minutes leading up to the phase of rapid acceleration. 

It has been realised for many years that the magnetic field configuration associated with
a sigmoid is highly likely to erupt \citep{canfield99}, 
however an exact observational definition of a sigmoid has not yet been reached. 
The term is used to describe continuous S-shaped structures
and also collections of loops that appear S-shaped overall. In the cases where a continuous
S-shaped sigmoid is observed, an eruption follows around within around 14 
hours \citep{green14}. Given the finding that continuous S-shaped sigmoids that form
along polarity inversion lines with sustained flux cancellation
are indicative of a low altitude flux rope, the data indicate that these flux ropes
can form and remain stable on the Sun for several hours.
Sigmoids are best observed near disk centre, so that projection effects do not
mask their shape. There is a related set of observations at the solar limb 
that indicate the formation
of flux ropes at a higher altitude, formed by coronal reconnection. These flux ropes
appear to form from around 7 hours to 10 minutes before their eruption 
\citep{patsourakos2013,cheng2011}. A recent study identified such a flux rope
near disk centre, allowing the photospheric magnetic field evolution to be observed
and revealing that the flux rope formation was driven by flux emergence which exhibited 
rotational motions of the fragments in the leading polarity, one about the other \citep{james2017}.
These photospheric motions and the corresponding evolution of the coronal field
led to magnetic reconnection in the corona and the formation of the flux rope.

As the time of the eruption approaches observations reveal that coronal 
structures begin to rise, albeit slowly, before the main phase of 
acceleration. Filaments darken and start their slow ascent
from around 3 hours before their successful eruption \citep{martin80}. 
More compact active region structures complete their slow rise phase more rapidly, 
taking as little as just 
a few minutes \citep[e.g.][]{zharkov11} before the structure undergoes rapid acceleration.

\subsubsection{Evolution of the eruption in the lower corona}
\label{subsec:evo_eruption}

Over many decades, observations of the onset and evolution of 
the eruption of a CME in the lower corona
have been drawn together to form the 
standard model for a CME or an eruptive flare. This model is also
known as the CSHKP model after the authors of the early papers 
\citep{carmichael64, sturrock66, hirayama74, kopp76}. 
For a review of some of the many
observational papers that 
have further supported this model see \cite{mckenzie02}.
The standard model of an eruption begins with the presence of a core field, either a sheared arcade or magnetic flux rope, 
embedded in an arcade field. Such sheared structures and flux ropes are well supported
by observations from the photospheric vector magnetic field to EUV and soft X-ray 
emission structures in the corona.
At the start of the eruption, the core field starts to rise,
stretching the overlying arcade. A current sheet forms in the arcade below the core field,
and reconnection sets in. Observations that support the occurrence of reconnection include
the reconfiguration of the coronal field, indications of non-thermal particles from the formation
of flare ribbons and hard X-ray emission, and reconnection inflow \citep{mckenzie02}.
These observational signatures indicate that reconnection is closely related in time to the rapid acceleration phase
of the CME. Indeed, the reconnection
helps facilitate the eruption as it builds poloidal flux around the erupting core field 
and cuts the tethers of the overlying field. Flare loops form below the 
reconnection region as the downward product of the reconnection.
Some CMEs show a clear rotation of the erupting structure as the eruption begins,
indicating the presence of a flux rope and a conversion of twist into writhe 
\cite[deformation of the axis of the flux rope,][]{green07}.

Not all CMEs that are initiated make it to a full eruption \citep{rust03}. 
Instead, their rapid ascension is arrested in the lower corona. This sub-set
of events are known as failed eruptions or confined eruptions. The failure of the eruption might be due to
the reconnection that occurs below the erupting core field not sufficiently cutting the tethers
of the overlying arcade field \citep{moore01} or the field strength in the overlying arcade
not dropping off sufficiently rapidly with height \citep{torok05}.
However, failed eruptions still serve a useful purpose in 
CME studies as they allow an investigation of the triggers versus the drivers of an eruption. 
In a failed eruption, presumably the trigger occurs, but the physical processes
needed to drive the CME into the heliosphere do not
(see Section \ref{sec:initiation}).

\subsection{Theory and simulations}
\label{sec:origin_theo_sim}

From a theoretical point of view, there is little doubt that large-scale solar eruptions are magnetically driven, since they originate in the corona where the plasma beta (ratio of gas to magnetic pressure) is small. They are powered by the free magnetic energy that is stored in the corona in non-potential magnetic fields, i.e., in volumetric electric currents. These currents are believed to emerge from the convection zone \cite[e.g.,][]{leka96} or be produced by photospheric flows that shear and twist the coronal magnetic field \cite[e.g.,][]{klimchuk92}. Eruptions occur when the slow, continuous stressing of the coronal field by flux emergence and surface flows reaches some threshold above which magnetic equilibrium cannot be maintained and the field violently erupts \citep{forbes2000}. While this overall picture is well established, the detailed processes that govern the evolution prior to and during an eruption are not yet fully understood. In this subsection, we thus focus on the following three questions: How is the free energy built up in the corona? (Sect.\,\ref{sec:build-up}) How are solar eruptions initiated? (Sect.\,\ref{sec:initiation}) What are the physical mechanisms that can propel the magnetic field and plasma into interplanetary space? (Sect.\,\ref{sec:background}) We will also discuss potential connections between individual eruptions (leading to ``sympathetic'' events) in Sect.\,\ref{ss:symp}.

\subsubsection{Modelling the Build-Up of Free Magnetic Energy} 
\label{sec:build-up} 

Non-potential magnetic fields are at the epicenter of solar eruptive behavior. Eruptions may originate from large-scale relatively weak magnetic fields as in the case of streamer blowouts or quiescent prominence eruptions \citep{hundhausen1993,lynch2010} or they may originate from strong active region magnetic fields. Whatever the circumstances, solar eruptions originate from strongly non-potential fields where the photospheric magnetic field is nearly parallel to the polarity inversion line (PIL), a so called magnetic shear \citep[e.g.][]{forbes2000,schrijver2009}, which stands in contrast to a potential field that runs perpendicular to the PIL and has no free energy.  There is enormous evidence for the existence of highly sheared magnetic fields associated with CMEs and large flares. At the photosphere, magnetic shear in CME-productive active regions is directly measurable with vector magnetographs 
\citep[e.g.][]{hagyard1984,zirin1993,falconer2002,yang2004,liu2005,sun2012} and the associated shear flows can be observed along the photospheric PIL \citep[e.g.][]{strous1996,yang2004,deng2006,sun2012}.

With such observations in mind, MHD simulations of solar eruptions have long invoked shear flows prescribed as boundary conditions to energize magnetic fields to produce large-scale solar eruptions. Many examples exist including simulations by \cite{steinolfson1991,mikic1994,wolfson1995,antiochos1999,amari2003,lynch2008,vanderholst2009}, which are successful in reproducing many characteristics of CMEs, filament eruptions and flares. 
CME-like eruptions have also been modelled with prescribed rotational flows that twist up the field \citep[e.g.,][see also Table\,\ref{t:mechanisms}]{amari96b,torok03,aulanier05}.
A significant limitation of these numerical models is that because of the enormous differences in plasma conditions, they all treat the corona as being disconnected from the solar interior. While this simplification greatly reduces computational costs, it also removes any possibility of elucidating the physical processes that by their very nature transport free magnetic energy and magnetic helicity from the solar interior to the corona. 
   
In contrast, simulations of magnetic flux emerging from the convection zone to the corona set the stage to understand how active region magnetic fields are organized and energized. Global-scale simulations of the solar convection zone have shown how magnetic fields are amplified by the dynamo to produce buoyant loops that capture active region characteristics such as Joy's law and sunspot asymmetries \citep[e.g.][]{spruit1982,abbett2001,fan2014,nelson2014}. However, constraints in timescale and numerical resolution prevent these simulations from reaching beyond 95\% of the solar radius. To incorporate the upper layers of the solar atmosphere, fully compressible MHD models are applied on small scales in Cartesian domains \citep[e.g.][]{shibata1989,matsumoto1993} that illustrate magnetic flux buoyantly rising through solar photosphere to the corona. Simulations that followed \citep{manchester2001,fan2001,magara2003,torok2014} found that under these circumstances, shear flows develop spontaneously where horizontal flows reverse direction across emerging loop structures. An analytical model by \cite{manchester2000} first showed that these shear flows are in response to the Lorentz force, which occurs as magnetic fields rise through the stratified atmosphere and are deformed by the intense pressure gradient of the surrounding plasma. The shearing motions are of the form of large-amplitude shear Alfv\'en waves where magnetic tension drives the flows that draw the magnetic field along the direction parallel to the polarity inversion line and transport magnetic flux and energy from the convection zone into the corona. \cite{manchester2003,manchester2004,manchester2007} found that such shear flows are capable of driving eruptions of magnetic arcades and emerging flux ropes and proposed this mechanism as an energy source and initiation mechanism for CMEs. 

%%------------------------------ FIGURE --------------------------------
\begin{figure*}[ht!]
\begin{center}
\includegraphics[width=1.\textwidth]{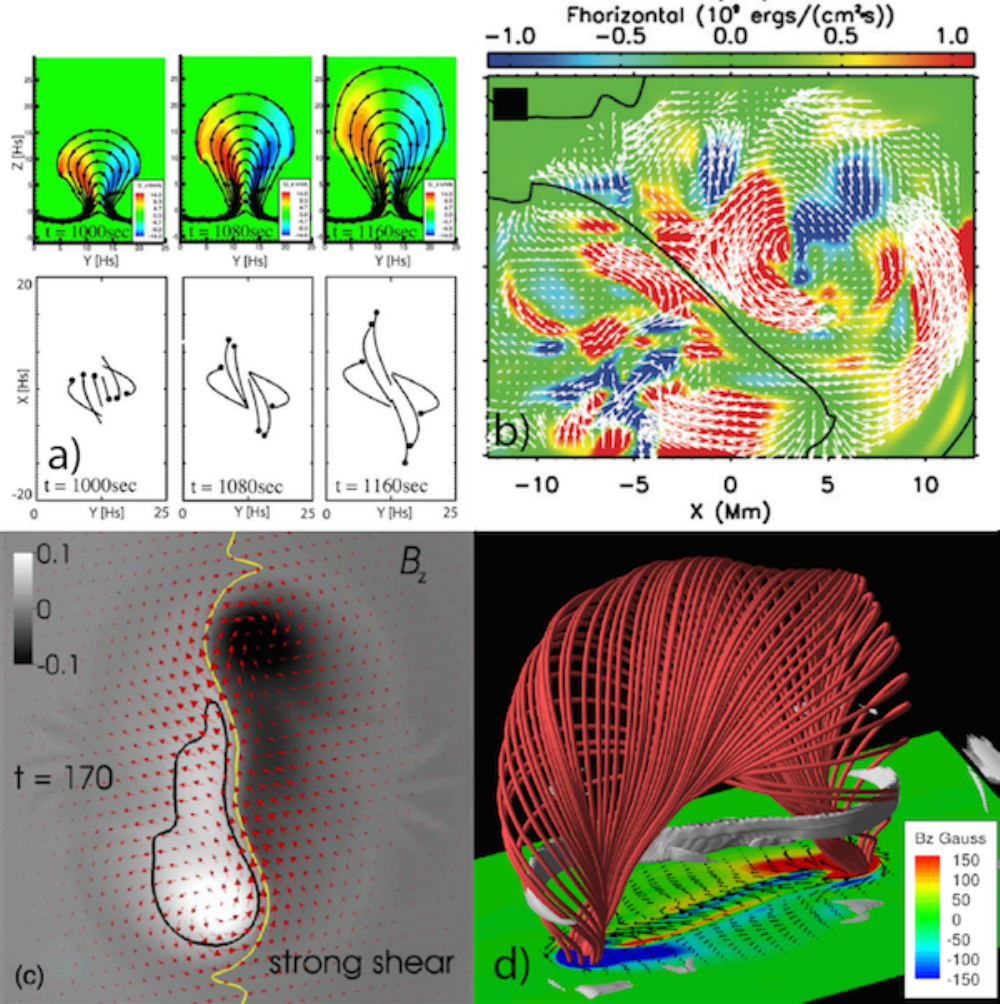}
\end{center}
\caption{
Simulations of shear and rotational flows, energy transport, and magnetic
flux-rope eruption. Panel (a) \citet[]{manchester2001} shows a 2.5D simulation
of flux emerging, with the top row showing the shear velocity shown over the cross 
section of the emerging loop, and the bottom row showing the legs of the loops 
where dots are at the foot points. Panel (b) \citet[]{fang2012b} shows the Poynting
flux associated with horizontal flows passing though the corona. White arrows 
show the velocity vectors. Panel (c) \cite{torok2014} shows late-phase signatures
of an emerged magnetic flux rope at the photosphere.  Red arrows show the horizontal
magnetic field direction while the yellow line shows the PIL. Note the magnetic 
shear with the alignment of the magnetic field with the PIL.
Panel (d) \citet[]{manchester2004} shows the eruption of the flux rope resulting 
from shear flows.  At the photosphere, the vertical magnetic field strength is 
shown in color and the horizontal direction is shown with vectors. Here, the current 
sheet is shown with a gray iso-surface, forming a sigmoid structure that runs 
nearly parallel to the highly sheared PIL.
} 
\label{eruptionx4}
\end{figure*}
%%-----------------------------------------------------------------------

Early simulations of flux emergence treated the solar interior as a simple polytropic layer, which roughly captured the temperature structure and density but was free of convection. In recent years, realistic solar models with granular convective and coronal atmospheres have been developed with the inclusion of radiative transfer and heat conduction. Examples include \cite{stein2006,abbett2007,rempel2009,nordlund2009,martinez2009,cheung2010,fang2010,kitiashvili2010}, which illustrate the complex and dynamic interaction between magnetic fields and granular convection. Work in this vein began with quiet sun magneto-convection \citep[e.g.][]{stein2006,abbett2007,rempel2014a}, which shows the development of a near-surface dynamo and formation of kiloGauss flux tubes in convective downdrafts. In related work, \cite{martinez2009,kitiashvili2010} found strong vortex flows in downdrafts at the vertices of convective granules where the twisting magnetic field produces strong outflows representative of spicules. The inclusion of strong large-scale vertical magnetic fields in the presence of convective flows leads to the formation of sunspots as shown by \cite{rempel2009,cheung2010,rempel2014b}.  

Following this work, came realistic simulations of flux emergence and active region formation in a fully convective layer \citep[e.g][]{fang2010,fang2012a,cheung2010,rempel2014b}. In the convection zone, high-plasma beta magnetic flux is severely distorted as distinct parts of the flux system rise and fall with the large-scale convective flows. As flux passes through the photosphere, it is shredded by the granular convection only to later reorganize as like polarities coalesce into kiloGauss pores and sunspots \citep[]{fang2012a,cheung2010,rempel2014b}. Even with this complex structure the Lorentz force continues to drive both shear and rotational flows that transport magnetic flux, helicity and free magnetic energy from the convection zone just as found in earlier simulations. In the case of \cite{fang2012b}, it was also found that while rotational flows provide free energy in the area of sunspots, shear flows dominate the energy build-up at the PIL. \cite{fang2012b} further found that converging motions, flux cancellation and tether-cutting reconnection can combine to build-up the magnetic shear and free energy in the corona necessary for eruptive events. In this case, nearly 40\% of the magnetic energy is free energy, which is near the amount required to produce eruptions \citep{amari2003,aulanier2010,moore2012}.

Recent observations are proving that the ubiquitous shear and rotational flows so strongly associated with CMEs produced during an active region’s emergence phase are driven by the Lorentz force. Analysis of high-resolution vector magnetograms \citep[]{georgoulis2012} indicate that photospheric electric currents generate the Lorentz force that drives shear flows along active region PILs during this phase. Similar analysis by \cite{su2008} has revealed that the Lorentz force is the driver of rotational flows in sunspots. In these cases, the flows are shown to produce a build-up of magnetic energy with magnetic stress passing from the convection zone into the corona to form non-potential fields. These observational results both verify prior numerical simulations and confirm a fundamental mechanism for energizing coronal magnetic fields before eruption. 

Simulation results of the development of shear flows, sheared magnetic fields, Poynting flux and an erupting flux rope are shown in Figure\,\ref{eruptionx4}. Panel (a) shows the results of a 2.5 D simulation of flux emergence that shows the development of sheared magnetic fields \citep[]{manchester2001}. Panel (b) shows the Poynting flux associated with horizontal flows passing into the corona \citep[]{fang2012b} while Panel (c) shows the development of a highly sheared magnetic field resulting from flux emergence \citep[]{torok2014}. Panel (d) shows the magnetic field of an erupting magnetic flux rope, where the eruption is triggered by shear flows driven by the Lorentz force.

\subsubsection{Initiation Mechanisms of Solar Eruptions}
\label{sec:initiation}

%%------------------------------ FIGURE --------------------------------
\begin{figure*}[t]
\begin{center}
\includegraphics[width=1.0\textwidth]{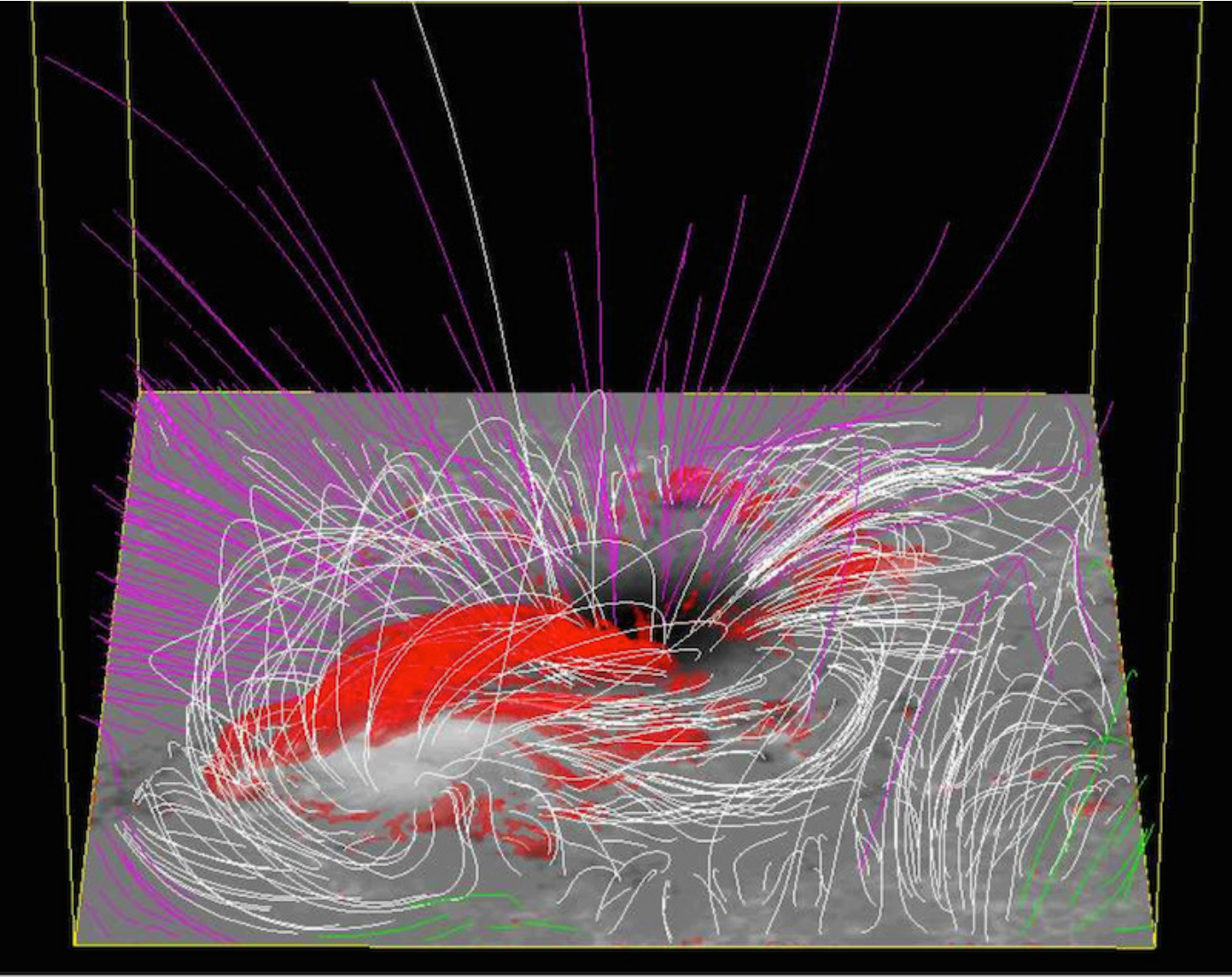}
\caption{
Non-linear force-free field extrapolation of NOAA AR 10930 before it produced an X3.4 flare and 
CME \citep[from][]{schrijver08a}. Red iso-surfaces show electric currents. Coloured 
(white) field lines are open (closed). Closed field lines arch over the currents, 
stabilizing the sheared or twisted core magnetic field.     
} 
\label{f:schrijver08}
\end{center}
\end{figure*}
%%-----------------------------------------------------------------------

As described in the previous sections, solar eruptions are powered by the free magnetic energy that is stored in current-carrying magnetic fields in the corona. Prior to an eruption, the magnetic configuration essentially consists of two parts. One part is the current-carrying (sheared and/or twisted) core field, which is always located above a polarity inversion line and which may or may not harbor a filament. This field naturally seeks to expand and must therefore be stabilized by the other part of the configuration, namely by the largely current-free (potential) ambient field that surrounds the core field (often referred to as the ``strapping field''), as illustrated in Fig.\,\ref{f:schrijver08}. The idea underlying many eruption-models is that an eruption can be initiated by any physical mechanism that is capable of bringing this configuration to a state at which stable equilibria cease to exist and the system violently erupts, thereby releasing some of the stored free energy. The initiation mechanism could be one that either increases the current in the core field or decreases the stabilizing magnetic tension of the ambient field, or both \citep[see, e.g.,][]{aulanier14}. 

However, even though eruptions have been observed extensively for many years, the specific build-up and trigger mechanisms,
and the mechanisms that drive CMEs remain rather elusive. This is, to a large extent, due to the persistent lack of routine magnetic field measurements in the corona. Analytical models and numerical simulations provide a useful tool for our attempts to close this gap. These approaches typically employ idealized initial configurations and a reduced set of the MHD equations, both of which are strong simplifications of reality. However, such a simplified approach has two significant advantages. On the one hand, it often allows one to isolate in the model the specific mechanism of interest. On the other hand, it keeps the computational costs moderate and thus provides the possibility of performing extensive parametric studies. Such studies are useful because they allow, for example, an assessment of physical quantities that are difficult or impossible to measure in the corona, such as magnetic twist \citep[e.g.,][]{kliem12}. In the past decades, idealized analytical models and numerical simulations have been widely used to test and study suggested eruption mechanisms, and have thereby greatly increased our understanding of how solar eruptions work. We note that significantly more complex MHD simulations of eruptions exist as well. These simulations include observed data and more of the physics, and they are aiming to reproduce specific eruptions with as much realism as presently possible (see Manchester et al. in this Volume).

%================================================================*
\begin{table}[!t]
\begin{center}
\begin{tabular}{V{3} p{2.77cm} V{3} p{7.87cm} V{3}}
\hlineB{3}
MECHANISM  &  REFERENCES  \\
\hlineB{3}
\vspace{-0.15cm}
\underline{TRIGGER:}      & \\
%Mass (un-)loading         & \cite{low96,fong02,zhang.m04,seaton11} \\
%\hline
Sunspot rotation          & \cite{amari96b,tokman2002,torok03,aulanier05,rachmeler2009} \\
\hline
Twisting overlying field  & \cite{torok2013} \\
\hline
Shearing of arcade        & \cite{mikic88,biskamp89,mikic1994,choe96a,choe96b,amari96a,jacobs06,jacobs09,roussev07,shiota08,downs11} \\
\hline
Reversed shear            & \cite{kusano04} \\
\hline
Self-induced shear flows  & \cite{manchester2003,manchester2007,manchester2004} \\ 
\hline
Magnetic breakout         & \cite{antiochos1999,macneice04,lynch04,lynch2008,vanderholst07,vanderholst2009,masson13} \\
\hline
Tether cutting            & \cite{moore92,moore01,aulanier2010} \\
\hline
Converging flows          & \multirow{2}{8.0cm}{\cite{inhester92,amari2003,roussev04,zuccarello12,mikic13}} \\
/ Flux cancellation       & \\
\hline
Flux decrease             & \multirow{2}{8.0cm}{\cite{lin.j98,linker01,linker03,amari00,amari03b,titov08,reeves10}} \\
/ dispersion              & \\
\hline
FE close/below flux rope  & \cite{chen00,lin.j01,shiota05,dubey06} \\
\hline
FE into potential arcade  & \cite{zuccarello08,jacobs12,roussev12} \\
\hline
FE into sheared arcade    & \cite{notoya07,kusano12} \\
\hline
Helical kink instability  & \cite{sakurai76,hood79,gerrard01,fan03,torok04,torok05} \\
\hline
Flux transfer/feeding     & \cite{zhang.q14,kliem14b} \\
\hline
Tilt instability          & \cite{keppens14} \\
\hline
Double-arc instability    & \cite{ishiguro2017} \\
\hlineB{3}
\vspace{-0.15cm}
\underline{DRIVER:}       & \\
Torus instability         & \multirow{4}{8.0cm} 
                          {\cite{vantend78,priest90,forbes91,lin.j98,kliem06,torok07,fan07,olmedo10,demoulin10,kliem14}} \\ 
/ Flux-rope catastrophe             & \\ 
/ Loss of equilibrium     & \\
                          & \\
                          \hline
Flare--reconnection       & \cite{lin.j00,vrs08angeo,temmer10,karpen12} \\
\hlineB{3}
\vspace{-0.15cm}
\underline{Review articles:} & \vspace{-0.15cm} \cite{forbes2000,chen.j01,klimchuk01,low01,priest02,lin.j03,linker03,zhang.m05,forbes06,moore06,mikic06,roussev08,vrs08angeo,amari09,linton09,schrijver2009,aulanier2010,chen.pf11,jacobs11,kleimann12,schmieder13,aulanier14,schmieder15,inoue16,chen.j17}\\
\hlineB{3}
\end{tabular}
\caption{Compilation of suggested physical mechanisms of the ``storage-and-release'' type for the triggering and driving of solar eruptions. ``FE'' stands for flux emergence. Note that (i) we do not distinguish here between 2D and 3D simulations; (ii) we do not include simulations that model CMEs by starting with an ``out-of-equilibrium'' flux-rope configuration \citep[e.g.,][]{roussev03,manchester04b,toth07,cohen09,lugaz11,pagano13}, even though they are compatible with the storage-and-release paradigm; and that (iii) each article is referenced only once, even if the model or scenario it describes involves more than one mechanism. The bottom row contains related review articles.}
\label{t:mechanisms}
\end{center}
\end{table}
%================================================================*

Table\,\ref{t:mechanisms} shows a compilation of eruption models of the so-called ``storage and release'' type, together with references and a compilation of review papers on the subject.\footnote{This table is not intended to be complete, more mechanisms and articles may be found in the literature. See \cite{aulanier14} for the differences between ``flux decrease'' and ``flux dispersion'' models.} Storage-and-release models assume that the free energy required to power an eruption is slowly (i.e., quasi-statically) accumulated in the corona over hours or days, and then rapidly released during the eruption. ``Directly driven'' models, such as thermal blasts during flares or flux injection from the solar interior, appear outdated and are not considered here (for discussions of their validity see, e.g., \citealt{forbes2000,klimchuk01,linker03}; but see also \citealt{chen.j17}). Furthermore, the table does not include eruption scenarios that have, to the best of our knowledge, not yet been tested using numerical simulations. Such scenarios are, for instance, the triggering of eruptions by the merging of magnetic flux ropes \citep[e.g.,][]{pevtsov96} or by the unloading of prominence mass \citep[e.g.,][]{low96,seaton11}, as well as the conjecture that CMEs occur in order to avoid an over-accumulation of magnetic helicity in the corona \citep[e.g.,][]{low94}.

The mechanisms are grouped according to their predominant role in the eruption process, i.e., whether they act as a ``trigger'' or as a ``driver''.\footnote{Any such grouping is somewhat subjective. \cite{aulanier14}, for example, considers magnetic breakout rather than flare-reconnection as the second driving mechanism next to the torus instability.} By driver we mean here any mechanism that can account for the rapid acceleration and huge expansion of plasma and magnetic field observed during eruptions, i.e., a mechanism that can produce a CME. As a trigger, on the other hand, we consider mechanisms that are capable of initiating an eruption, but can by themselves not produce a CME. The role of a trigger is rather to bring the system to a state at which a driver can take over and complete the eruption. 

Such a strict distinction between trigger and driver is somewhat artificial, but we use it here for the following reasons. First, rise profiles of eruptions sometimes clearly show two distinct phases that cannot be fitted with one functional form \citep[see Section\,\ref{subsec:shortterm_obs} and e.g.,][]{sterling07a,sterling07b,schrijver08b}, indicating the occurrence of two distinct physical mechanisms. Second, it has been demonstrated in MHD simulations that certain mechanisms (such as tether-cutting and magnetic breakout) do not seem sufficient for producing a full eruption \citep[e.g.,][]{aulanier2010,karpen12}. Finally, some mechanisms do not appear to be capable of producing a CME, just by their nature. The helical kink instability, for example, is expected to lead to a (relatively moderate) deformation of the shape of an unstable flux rope, but not to its huge expansion.    

We note that some of the processes listed as a trigger mechanism in Table\,\ref{t:mechanisms} (flux emergence, sunspot rotation, flux cancellation) typically act on time-scales of days, much longer than the typical duration of the slow rise phase preceding eruptions. Their preliminary role is to accumulate free energy in the corona (see Sect.\,\ref{sec:build-up}). However, as demonstrated by the numerical simulations referenced in the table, they can also, on shorter time scales, induce changes to the system that push it towards an eruption (an example would be flux emergence into an existing pre-eruptive configuration). In that sense it appears justified to consider these processes not just as build-up, but also as trigger mechanisms.

At present, it seems that there may exist only two driving mechanisms in solar eruptions, namely the torus instability and the ``flare reconnection'' \citep[for a somewhat different view, see][]{aulanier14}. The latter is the reconnection that takes place across the vertical current sheet below a CME and produces the flare signatures. The torus instability is an ideal MHD instability, closely related to the processes known as ``flux-rope catastrophe'' and ``loss of equilibrium'' (see \citealt{demoulin10} and \citealt{kliem14} for a detailed discussion). In this instability, the driver responsible for the acceleration of the ejecta is the hoop force or Lorentz-self force, which results from the curvature of a current-carrying flux rope \citep[e.g., ][]{titov99}. The flare reconnection, on the other hand, can drive the eruption in several ways. For instance, by transferring restraining overlying flux into flux of the erupting sheared core or flux rope, or by producing highly bent field lines in a rope that accelerates it by means of a slingshot effect \citep[e.g.,][]{linton01,mackay06,archontis2008}. These two processes (instability and reconnection) often occur simultaneously and are closely coupled \citep[e.g.,][see also Sect.\,\ref{sec:background}]{bein12,vrs16}. Their respective contributions to the acceleration of the ejecta are presently not well known; they likely depend on various parameters and may therefore differ from case to case. Disentangling the contributions is not trivial and poses a challenge for modellers. As we will see in the next section, it appears that at least the most impulsive eruptions are driven predominantly by reconnection.  

As apparent from Table\,\ref{t:mechanisms}, many more trigger mechanism than driving mechanisms seem to exist. Rather than describing all the trigger mechanisms that have been suggested, we only give a few general remarks here, noting that more detailed explanations can be found in the references compiled in Table\,\ref{t:mechanisms}. First of all, 
trigger mechanisms should not be thought of as acting in isolation or being mutually exclusive. It is likely that several of them work together, simultaneously or successively, in starting an eruption \cite[e.g.,][]{williams05}. Unfortunately it is often difficult, if not impossible, to pin down which particular mechanism or combination of mechanisms triggered a specific eruption. Most mechanisms cannot be observed directly, so indirect evidence such as EUV brightening or the slow rise of a filament must be used. Also, the same observational manifestations may be produced by different mechanisms. This lack of detailed knowledge greatly hampers our capabilities to forecast the onset of eruptions (see Sect.\,\ref{sec:predict}), as different trigger mechanisms may have very different onset conditions. Moreover, even if one could predict that a certain trigger mechanism will take place, it would still be hard to predict whether an eruption will produce a CME or remains confined \citep[e.g.,][]{moore01,torok05,guo.y10} as we know little about the conditions under which trigger mechanisms can succeed in switching on driving mechanisms. In fact, it seems that, at least in some events, several confined eruptions are required before a full eruption can occur \citep[e.g.,][]{panesar15,chintzoglou15,liu.r16}. 

Overall, observations and simulations have helped us to develop a relatively clear qualitative picture of how a typical solar eruption works. Slow magnetic reconnection acting below (potentially also above) a sheared or twisted core field results in a slow rise of this field. During this slow rise, the core field successively detaches from the photosphere, reconnection occurs at increasingly larger heights in the corona, and the field successively transforms from a structure that was initially probably more like an arcade into a flux rope. Once the rope has been lifted to a sufficiently large height, the torus instability sets in and starts to rapidly accelerate the rope upward. More or less simultaneously, the flare reconnection sets in and supports the rapid acceleration. What is still needed, though, is a better understanding of the onset conditions, efficiency, and interplay of the underlying mechanisms. In the next sub-section, we will use some general considerations and an analytical approach to dig deeper into the physics behind driving mechanisms and the acceleration of CMEs.

\subsubsection{Eruption driving mechanisms and physical background}
\label{sec:background}

The free magnetic energy contained in the coronal field can be expressed in the form $W=LI^2$, where $I$ is the total electric current flowing through the system and $L$ is the self-inductance that includes information on the entire magnetic-field geometry \citep[e.g.,][and references therein]{garren94,chen89,chen03,jackson04,zic07}. Depending on the desired complexity of the model, an otherwise intricate coronal magnetic configuration is represented in the analytical approach as a simple line current, a flux rope, a flux rope embedded in a magnetic arcade, etc. In the process of an eruption, the free magnetic energy is explosively released
and transformed into kinetic and gravitational potential energy. In addition, if the eruption is accompanied by a flare, part of the energy is consumed for various thermal and non-thermal processes (heating, convective flows, non-thermal particle populations, etc.). Finally, as the speed of the eruption increases, progressively more and more energy is spent for the work against the ``aerodynamic'' drag. Let's now go on to consider the maximum velocity and acceleration that a CME can attain.

Bearing in mind energy conservation, one can relate the kinetic energy density and the magnetic field energy density as $\rho v^2$/2 $< B^2$/2$\mu$, where $\rho$, $v$, $B$, and $\mu$ are the plasma mass-density, flow speed, magnetic field strength, and permeability, respectively. This relation can be rewritten as $v<B/\sqrt{\rho\mu}$, i.e., $v<v_{\rm A}$, where $v_{\rm A}$ is the Alfv\'en speed \emph{within} the erupting structure.
Consequently, it is expected that eruptions from active regions should attain higher speeds than those occurring in quiet-sun regions, which is fully consistent with observations  \citep[e.g.,][and references therein]{vrs05,bein11,bein12}.

The CME acceleration is driven by the Lorentz force, or formulated in a more simplified way, by the gradient of the magnetic pressure $\nabla (B^2/2\mu)$. Taking an order of magnitude estimate for the magnetic pressure as $\nabla (B^2/2\mu)\approx B^2/2d\mu \approx f_L$, where $d$ is the size of the structure and $f_L$ is the Lorentz force \emph{per} unit volume, the acceleration can be expressed as $a=f_L/\rho=B^2/2d\rho\mu$, i.e., $a\approx v_{\rm A}^2/2d$. This implies that the acceleration of a CME is limited by the intrinsic Alfv\'en speed, and that the acceleration should be larger in compact eruptions than in extended ones. Indeed, in eruptions of active-region structures, where the Alfv\'en speed is high and the source region is rather compact, the acceleration  is usually much larger than in eruptions from quiet-sun regions \citep[e.g. eruptions of quiescent prominences, stealth CMEs; see, e.g.,][]{vrs05,bein12,howard13}. Note that the relation
$a\approx v_{\rm A}^2/2d$ implies that in the case of very compact sources, $d<100$\,Mm, accelerations can achieve values on the order of 10\,km\,s$^{-2}$, such as is observed in the most impulsive events \citep{vrs07,bein11}.

Using these estimates for the maximum velocity and acceleration, one can also roughly estimate the time and distance over which an eruption will be accelerated. The acceleration time can be expressed as $t_a\approx v/a$, implying $t_a\approx d/v_{\rm A} = t_A$, where $t_A$ represents the Alfv\'en travel time over the eruptive structure. The acceleration, $a$, speed, $v$, and the acceleration distance, $d_a$, can be related as $a\approx \sqrt{2ad_a}$, implying $d_a\approx v^2/2a=d$, i.e., the acceleration distance should be comparable with the size of the eruptive structure. Indeed, both types of relationships $t_a\propto d$ and $d_a\propto d$, as well as $t_a\propto1/a$, are found in observations \citep[][]{vrs07,bein11}. Furthermore, it is well known that eruptions originating from active regions (smaller size, higher Alfv\'en speed) are more impulsive (stronger acceleration over a shorter acceleration time) than those launched from quiet-sun regions \citep[][]{zhang06,vrs07,bein11}, consistent with $t_a\approx d/v_{\rm A}$ and $a\approx v_{\rm A}^2/2d$.
Characteristic values for active-region and quiescent-prominence eruptions, based on the presented order-of-magnitude relations, are displayed in Table~\ref{estimates}.

\begin{table}[htbp]
\center
  \caption{Typical values for active region (AR) and quiescent prominence (QP) eruptions. The size of the CME source structure is given by $d$, the Alfv\'en speed within the erupting structure by $v_{\rm A}$, velocity by $v$, acceleration by $a$, acceleration duration by $t_a$ and distance over which the structure is accelerated by $d_a$.}
     \begin{tabular}{ccccccc}
\hline
       & $d$ [Mm]  & $v_{\rm A}$ [km\,s$^{-1}$]   &  $v$ [km\,s$^{-1}$]   & $a$ [m\,s$^{-2}$]  & $t_a$ [s]  & $d_a$ [Mm] \\
\hline
AR   &    100     &      1000    &    1000   &     10000   &     100    &      100  \\
%\hline
QP   &   1000     &      400     &    400     &       160     &    2500   &    1000  \\
\hline
    \end{tabular}
  \label{estimates}
\end{table}

Let us now consider some general aspects related to the evolution of an eruption. The self-inductance $L$ is primarily dependent on the size $d$ of the current-carrying structure, $L\propto d$ \citep[][]{jackson04}. Thus, as an erupting structure expands, the self-inductance increases, and it can be approximately taken that $d$ is proportional to the heliospheric distance $r$, i.e., $d\propto r$. Under this assumption $L\propto r$. On the other hand, the self-inductance of the magnetic structure is by definition related to the electric current as $\Phi=LI$, where $\Phi$ is the magnetic flux associated with the current $I$  \citep[see, e.g.,][]{B&T62,zic07}. In the absence of dissipative processes that could lead to magnetic reconnection, the flux $\Phi$ stays preserved, implying that the electric current has to decrease as $I\propto1/r$. This means that the free magnetic energy of the system, $W=LI^2$, decreases, being transformed into the kinetic energy and the work done against drag and gravity.

In line with the magnetic origin of an eruption, the CME acceleration is driven by the Lorentz force, or formulated in a simplified way, by the Amp\`ere force {\it\textbf{F}}\,=\,{\it\textbf{I}}\,$\times${\it\textbf{B}}. Consequently, in the absence of reconnection, the driving force of a CME rapidly decreases with distance from the Sun, because both the electric current $I$ and the magnetic field $B$ protruding through the magnetic structure decrease \citep[for the dependence of the magnetic field on height see, e.g.,][]{dulk78,gary01,vrs02,vrs04bsplitIII,liu08}. Indeed, observations indicate that the driving force decreases approximately as $r^{-\alpha}$, where the exponent ranges between $\alpha\approx1$ and 2 \citep[][]{vrs01,vrs04domi,vrs06asr}. In other words, the inductive effects and the decrease of the magnetic field with height, act to prevent a very strong acceleration happening over a large time and distance range.

The general physical principles that govern the eruption process described here should be incorporated in any quantitative model,
no matter what the specific magnetic configuration is. Most often, the eruptive structure is represented by a semi-toroidal magnetic flux rope that is fixed at both ends in the photosphere. It is worth noting that the flux-rope concept has been in use since before the discovery of CMEs, i.e., in early models of solar flares and eruptive prominences \citep[e.g.,][]{gold60,ellison61,gold62,byrne64,hyder67,anzer70}. Gradually, flux-rope models developed from very simple ones \citep[e.g.,][]{K&R74,mouschovias78,sakurai76,anzer78,vantend78,anzer82,pneuman80,vrs84,steele89} to more complex forms, where various relevant effects have been sequentially taken into account \citep[e.g.,][]{chen89,vrs90,titov99,torok03}. In addition, the development of numerical techniques has provided more demanding studies, beyond those provided by the analytical approach.

%%%%%%%%%%%%%%%%%%%%%%%%%%%%%%%%%%%%%%%%
    \begin{figure}
    \center
  \includegraphics[width=1.\textwidth]{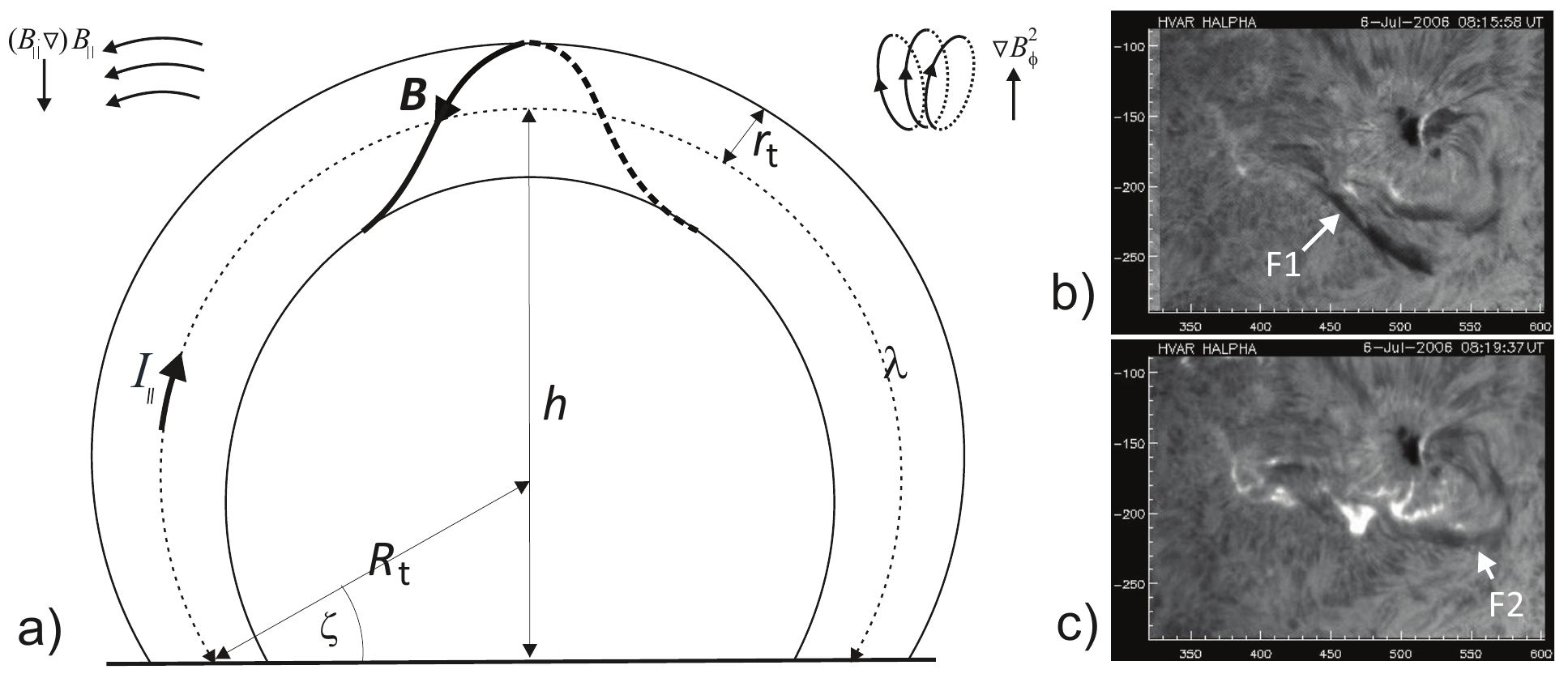}
\caption{a) Cartoon of a semi-toroidal flux rope with definition of symbols used in the text. In the upper-left and upper-rigth corner the effects of magnetic tension and pressure gradient are sketched, respectively. b) \& c) Eruption of July 6, 2016: Filament F1 in a slow-rise phase (b), indicating a presence of a metastable flux rope, gradually evolving towards the loss-of-equilibrium state;  The eruption of the filament F1 caused a flare below it (c) and destabilized the nearby (also metastable) filament F2 that erupted too, causing a westwards extension of the flare (Hvar Observatory filtergrams; x- and y-coordinates are shown in arcsec).}
\label{fig:rope}
\end{figure}
%%%%%%%%%%%%%%%%%%%%%%%%%%%%%%%%%%%%%%%%

Now we focus on the implications of the basic physical principles that have just been discussed as applied to the flux rope concept of an eruption.
To do so, let us consider the forces acting on a semi-toroidal magnetic flux rope, sketched in Fig.~\ref{fig:rope}a.
In such a configuration, several effects have to be considered. First, the curvature of the flux rope axis causes the Lorentz self-force (also known as ``hoop force''). The contribution of the self-force to the equation of motion was firstly estimated by applying the ``virtual work'' principle \citep[e.g.,][]{shafr66,garren94}.
Hereafter, we will follow the procedure proposed by \cite{vrs84} and \cite{vrs90}, where the self-force is estimated considering directly the Lorentz force components related to the magnetic pressure gradient and tension (sketched in the upper right and left corners of Fig.~\ref{fig:rope}a).   

Next, one has to take into account a diamagnetic effect, which comes from the influence of eddy electric currents induced at the solar ``surface''.  
In the simplest way, this effect can be roughly estimated by considering the effect of a line-current located at the height $h$ above the conducting surface, as done by \citet{K&R74}. Furthermore, the flux rope concept includes an overlying coronal field ($B_{\rm c}$), and the influence of this field must be taken into account. The effects depends on the orientation of the background field -- in the case of the Kuperus-Raadu configuration \citep{K&R74}, the force is directed downwards, whereas in the case of the Kippenhahn-Schl\"uter configuration \citep{KS57}, the force is directed upwards.

Finally, one has to consider the effect of gravity. Taking into account buoyancy, the acceleration reads $ a_{\rm g}= -g (\rho-\rho_{\rm w})/\rho$,
where $g$ is the local gravitational acceleration and $\rho$ and $\rho_{\rm w}$ represent the density of the CME and the ambient plasma, respectively. Note that in the case $\rho\approx\rho_{\rm w}$ the gravitation can be neglected, 
and if $\rho<\rho_{\rm w}$ then $a_{\rm g}$ becomes positive (i.e., acts upwards) \citep{low82}.

Let us now consider the stability of a flux-rope and the conditions under which it will erupt.
Putting all the previously mentioned effects together leads to an expression for the net acceleration (for details of the derivation see \citealt{vrs08angeo} and \citealt{vrs16}): 
%%%%%
     \begin{equation}
      a= \frac{C_{\rm L}}{\Lambda}\left[\frac{1}{2\tilde{R}_{\rm t}} - \frac{1}{\tilde{R}_{\rm t} X^2} + \frac{1}{H}\right] - \frac{C_{\rm c}}{\Lambda^2\tilde{R}_{\rm t}} - C_{\rm g}\,g \,,
      \label{Eq:net2}
      \end{equation}
 %%%%%%
where $\tilde{R}_{\rm t}$, $\Lambda$, and $H$ represent the major radius of the semi-toroidal flux rope, the length of its axis, and the height of the axis summit, respectively, all normalized with respect to the footpoint half-separation (following the notation presented in Fig.~\ref{fig:rope}a, one can write $\tilde{R}_{\rm t}\equiv R_{\rm t}/d$, $\Lambda\equiv\lambda/d$, $H\equiv h/d$, and $\tilde{r}_{\rm t}\equiv r_{\rm t}/d$). The parameter $X$ represents the ratio of the poloidal and longitudinal magnetic field components, $X=B_{\phi}/B_{\parallel}$, at the surface of the flux rope (Fig.~\ref{fig:rope}a).
The first three terms on the right-hand side of Equation~(\ref{Eq:net2}), written within the square brackets, represent the effects of the magnetic pressure gradient, magnetic tension, and diamagnetic effect, respectively. The fourth term is due to the background field, whereas the last term considers gravity. 

In Equation~(\ref{Eq:net2}), we took into account the previously inferred relation $I\propto 1/\lambda$, i.e., we express the current as $I=I_0\lambda_0/\lambda$, where $I_0$ is the value of the axial current when the flux-rope axis is at the height $h=d$ (i.e., $R_{\rm t}=h=d$, $\lambda=R_{\rm t}\pi$, $\tilde{R}_{\rm t}=H=1$, $\Lambda=\pi$) and we considered that the mass of the flux rope is $M=const$. The constant $C_{\rm L}=\mu I_0^2\pi/4M$ determines the overall ``strength'' of the eruption, and it shows that the CME acceleration is proportional to $I_0^2$. Taking into account typical values of $M=10^{12}$\,--\,$10^{13}$\,kg \citep[e.g.,][]{vourlidas10,webb12} and $I_0=10^{10}$\,--\,$10^{11}$\,A \citep[e.g.,][and references therein]{hofmann91,Filippov15},
it can be estimated that the value of $C_{\rm L}$ should typically range between 100 and 1000 m\,s$^{-2}$.

Furthermore, we assumed that $B_{\rm c}$ (background coronal magnetic field surrounding the rope) is uniform over the rope. So the flux conservation $B_{\rm c}\,r_{\rm t}\lambda=const.$, implies $B_{\rm c}=B_{\rm c0}r_{\rm 0t}\lambda_0/r_{\rm t}\lambda$. We note that in the absence of reconnection the eruptive structure should expand in a self-similar way, i.e., $r_{\rm t}\propto R_{\rm t}$ \citep{vrs08angeo},
which means $B_{\rm c}=B_{\rm c0} R_{\rm 0t}\lambda_0 /R_{\rm t}\lambda = B_{\rm c0}\pi/\tilde{R}_{\rm t}\Lambda$. Furthermore, we introduced a constant $C_{\rm c}=I_0B_{\rm c0}\pi^3d/M$, where we have taken into account $I\lambda=I_0\lambda_0$, i.e., $I=I_0\pi/\Lambda$. Taking typical values of $d=100$\,Mm and $B_{\rm c0}=0.1$\,--\,1 G (i.e., $10^{-5}$\,--\,$10^{-4}$\,T), together with the previously used values for $I_0$ and $M$, one finds $C_{\rm c}$ could be as large as $5\times 10^{4}$ m\,s$^{-2}$. In the case where the effect of a decrease of the background magnetic field with height is considered, $B_{\rm c0}$ should be replaced by, e.g., $B_{\rm c}=B_{\rm c0}(R_{\rm 0t}/R_{\rm t})^n=B_{\rm c0}\tilde{R}_{\rm t}^{-n}$, as is often used in treatments of the torus instability \citep[e.g.,][and references therein]{kliem06,demoulin10,olmedo10}. Under this approximation, the last term in Equation~(\ref{Eq:net2}) should be replaced by $C_{\rm c}/\Lambda^2\tilde{R}_{\rm t}^{1+n}$.

To solve Equation~(\ref{Eq:net2}), one has to specify the relationship between the geometrical parameters $\tilde{R}_{\rm t}$, $H$, and $\Lambda$. Using Fig.~\ref{fig:rope}a, where we have included the angle $\zeta$ that represents the inclination of a line connecting a flux-rope footpoint and the center of curvature of the rope axis, one finds the parametric relationships $\tilde{R}_{\rm t}= 1/{\rm cos}\zeta$, $H = (1+{\rm sin}\zeta)\,/{\rm cos}\zeta$, and $\Lambda = (\pi+2\zeta)\,/{\rm cos}\zeta$.
Taking into account that the poloidal-to-axial field ratio can be expressed as $X\equiv B_{\phi}/B_{\parallel} = 2\,\pi\,\tilde{r}_{\rm t}/\Lambda$, Equation~(\ref{Eq:net2}) constitutes an explicit expression for the acceleration as a function of the angle $\zeta$, and from that, one can establish a functional dependence $a(H)$.

 %%%%%%%%%%%%%%%%%%%%%%%%%%%%%%%%%%%%%%%%
    \begin{figure}
    \center
  \includegraphics[width=1.\textwidth]{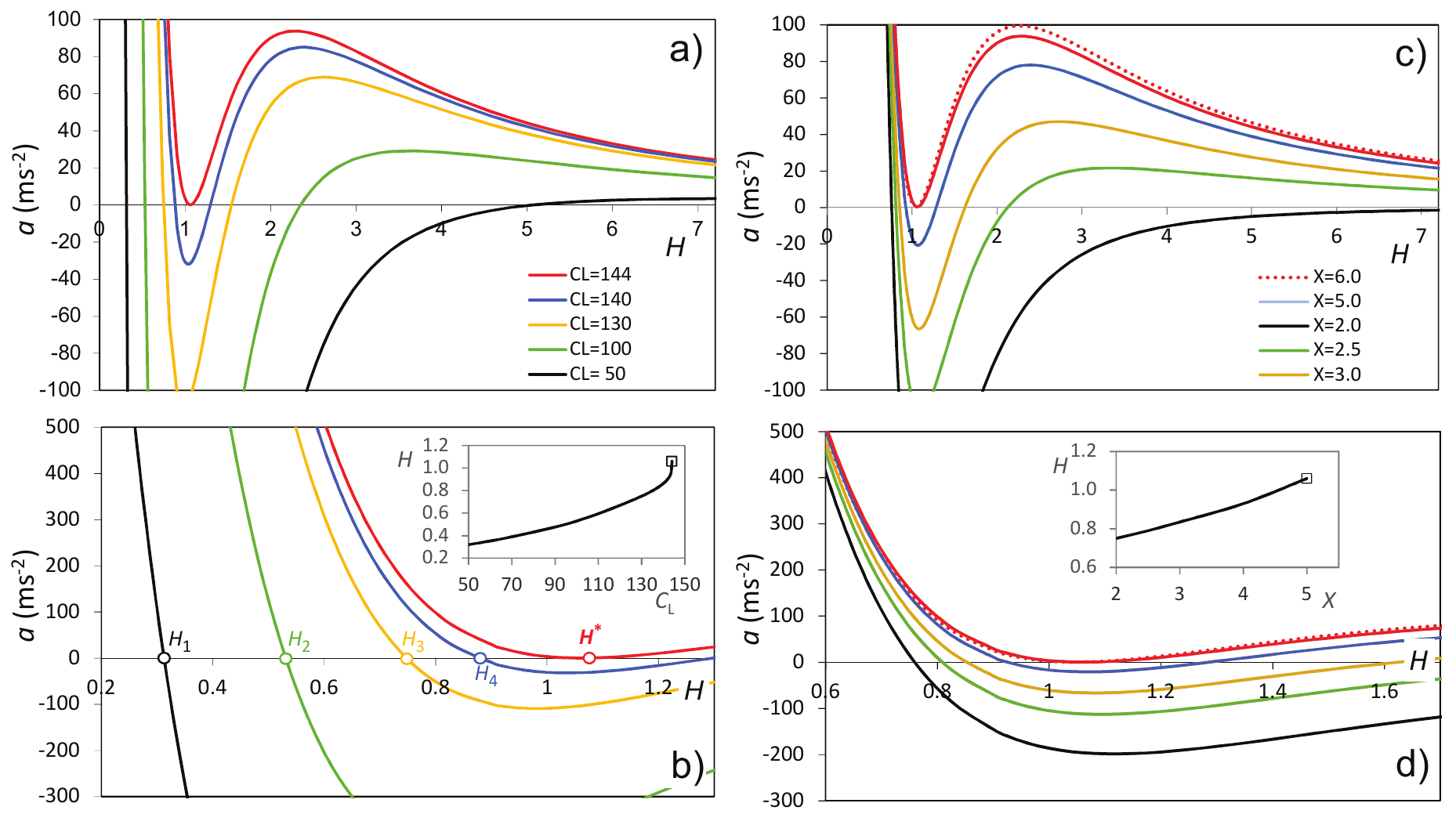}
\caption{ Various solutions of Equation~(\ref{Eq:net2}), representing stable, metastable, and unstable ropes.
a) $a(H)$ for various values of $C_{\rm L}$ (written in the legend), in combination with $X=5$, $r_{\rm t}/R_{\rm t}=0.1$, $C_{\rm c}=1.3\times10^{4}$ m\,s$^{-2}$.
b)  Enlarged part of a) around equilibrium points; in the inset the equilibrium height $H_e$ is shown as a function of $C_{\rm L}$.
c) $a(H)$ for various values of $X$ (written in the legend), in combination with $C_{\rm L}=144$\,m\,s$^{-2}$, $r_{\rm t}/R_{\rm t}=0.1$, $C_{\rm c}=1.3\times10^{4}$\,m\,s$^{-2}$; the red-dotted curve represents the case of $X=6.0$ with $C_{\rm L}=143$\,m\,s$^{-2}$.
d) Enlarged part of c) around equilibrium points; in the inset the equilibrium height $H_e$ is shown as a function of $X$.  }
\label{fig:evolution}
\end{figure}
%%%%%%%%%%%%%%%%%%%%%%%%%%%%%%%%%%%%%%%%

Various solutions to Equation~(\ref{Eq:net2}) are given in  Fig.~\ref{fig:evolution}.
Several $a(H)$ curves are shown, following an increasing $C_{\rm L}$ (Fig.~\ref{fig:evolution}a and b)  and $X$ (Fig.~\ref{fig:evolution}c and d) sequence. 
The increase of $C_{\rm L}$ and/or $X$ can be caused by various evolutionary processes related to the photospheric and subphotospheric motions.  For example, twisting motions at the flux-rope footpoints would lead to a transport of poloidal flux into the rope (thus increasing $X$) and increase the electric current (thus increasing $C_{\rm L}$). A similar effect may also be expected in the case of reconnection below the rope, which tends to increase the poloidal flux in the rope \citep[see e.g.,][]{vrs16}. The value of $C_{\rm L}$ could also be increased by emerging flux, since the increase of the total magnetic flux encircled by the current circuit associated with the rope would lead to an enhancement of the current. Finally, $C_{\rm L}$ might be increased by a mass leakage through the flux-rope legs, since $C_{\rm L}\propto 1/M$.

The intercepts of the $a(H)$ curves with the x-axis are equilibrium points ($a=0$). If the slope of the $a(H)$ curve, $\partial a/\partial H\equiv\omega^2$, at the equilibrium point is characterized by $\omega^2<0$ the equilibrium is stable (points denoted in Fig.~\ref{fig:evolution}b as $H_1$, $H_2$, $H_3$, and $H_4$). If the rope is moved from this position, it will oscillate around it, with a period of $P=2\pi/\omega$. Note that from the observational point of view, oscillations of coronal structures are especially well visualized in the case of prominence oscillations initiated by disturbances coming from distant eruptions, a phenomenon being reported as ``winking filaments'' a half of century ago by \citet{ramsey66}. Inspecting Figs.~\ref{fig:evolution}b and \ref{fig:evolution}d, one finds that as the value of $C_{\rm L}$ or $X$ increases, the equilibrium position shifts to larger heights (see the insets in Figs.~\ref{fig:evolution}b and \ref{fig:evolution}d) and the oscillation period increases (the slope $\partial a/\partial H$ becomes less steep).

The x-axis intercepts of $a(H)$ curves characterized by $\partial a/\partial H>0$ represent unstable equilibria. If the rope is displaced from a stable equilibrium to an unstable one (e.g., by a disturbance from a distant eruption), it will erupt. Thus, ropes that are characterized by $a(H)$ curves that have both types of x-axis intercepts are in a metastable state. Note that with increasing $C_{\rm L}$ or $X$, the distance between stable and unstable equilibria decreases, i.e., it becomes easier to push the rope from the stable to the unstable point and trigger the eruption. Finally, it should be noted that the $a(H)$ curve achieves the metastable form only for a sufficiently large value of $X$. This is illustrated in Fig.~\ref{fig:evolution}d, where the $a(H)$ curve for $X=2$ (black curve) is shown, which is characterized by only one (stable) equilibrium point. The curve calculated for $X=2.5$ (green curve) already contains the unstable equilibrium at $H\approx2$. Thus, a value of $X$ at which the magnetic structure transforms from stable to metastable configuration lies between 2 and 2.5; a more detailed analysis shows that for the considered combination of parameters ($C_{\rm L}=144$\,m\,s$^{-2}$, $r_{\rm t}/R_{\rm t}=0.1$, $C_{\rm c}=1.3\times10^{4}$\,m\,s$^{-2}$) this transition occurs at $X=2.2$.

The increase of $C_{\rm L}$ or $X$ leads to another essential physical situation. As the $a(H)$ curves gradually shift ``upwards'', at a given critical value of $C_{\rm L}^*$ or $X^*$, the stable and unstable equilibrium points merge (see the red curve in Fig.~\ref{fig:evolution}), after which there is no equilibrium possible. The local minimum in the $a(H)$ curve just touches the x-axis (in Fig.~\ref{fig:evolution}b this critical height is denoted as $H^*$), i.e., the whole $a(H)$ curve is located in the $a\ge0$ region. Consequently, the rope erupts, dynamically trying to find a new equilibrium. In the case when the gravity is neglected ($C_{\rm g}=0$) a new equilibrium does not exist, and the rope erupts. Note that the critical value $C_{\rm L}^*$ is lower for a higher value of $X$. This is illustrated in Fig.~\ref{fig:evolution}c, where the loss-of-equilibrium curve for $X=5$ (full-red line; $C_{\rm L}^*=144$\,m\,s$^{-2}$) is compared with the analogous curve for $X=6$ (dotted-red line; $C_{\rm L}^*=143$\,m\,s$^{-2}$). An example of a metastable flux rope containing filament plasma, gradually evolving towards the loss-of-equilibrium stage and eruption, is presented in Fig.~\ref{fig:rope}b. Filament F1 was slowly rising until reaching a critical height, after which it erupted (Fig.~\ref{fig:rope}c). Besides causing a flare below it, the eruption destabilized and triggered the eruption of the nearby metastable flux rope that was embedding the filament F2 \citep[for details see][]{miklenic09}.

 %%%%%%%%%%%%%%%%%%%%%%%%%%%%%%%%%%%%%%%%
    \begin{figure}
    \center
  \includegraphics[width=1.\textwidth]{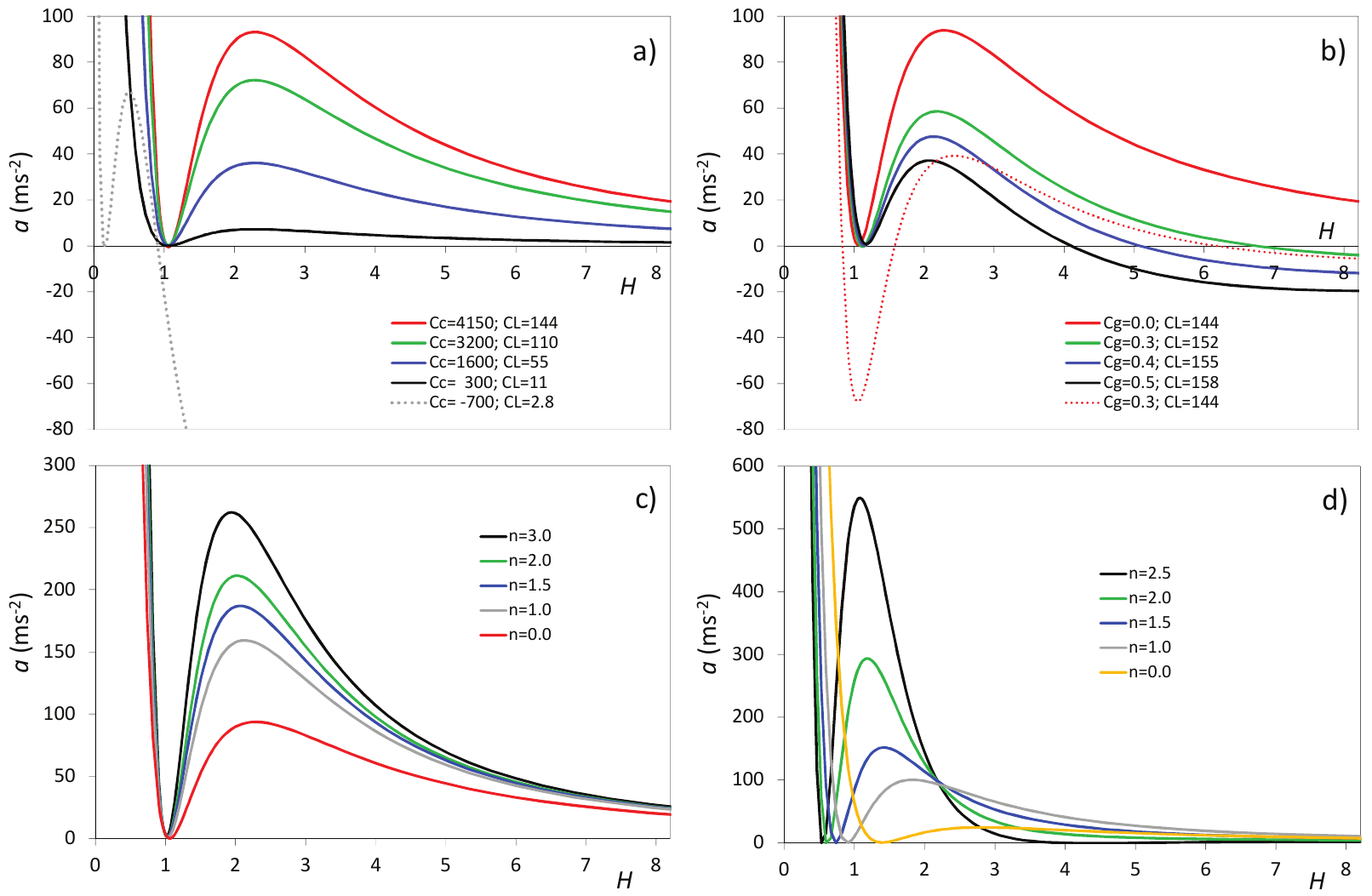}
\caption{ Flux ropes in the loss-of-equilibrium state: a) $a(H)$ solutions for different levels of the background-field effect (values of $C_{\rm c}$ expressed in m\,s$^{-2}$ are written in the legend, together with the corresponding critical values of $C_{\rm L}^*$); gray-dotted line represents Kippenhahn-Schl\"uter configuration with $C_{\rm g}=1$ ($IB_{\rm c}<0$, i.e., $C_{\rm c}<0$).
b) The gravity effect (values of $C_{\rm g}$ are written in the legend, together with the corresponding critical values of $C_{\rm L}^*$) calculated using $d=70$\,Mm, $X=5$, and $C_{\rm c}=13000$\,m\,s$^{-2}$. c) The effect of decreasing background field (values of the decay index $n$ are given in the legend), calculated using $X=5$, $C_{\rm L}^*=144$\,m\,s$^{-2}$, $C_{\rm c}=13000$\,m\,s$^{-2}$, and , $C_{\rm g}=0$. d) The torus-instability effect (values of $n$ are given in the legend), calculated using the same set of parameters as in c).
}
\label{fig:peakacc}
\end{figure}
%%%%%%%%%%%%%%%%%%%%%%%%%%%%%%%%%%%%%%%%

For a flux rope characterized by a given value of the poloidal-to-axial field ratio $X$, a critical value $C_{\rm L}^*$ at which the eruption starts depends on the relative contribution of the background magnetic field and the gravity. In other words, it depends on the values of $C_{\rm c}$ and $C_{\rm g}$, which is illustrated in Figs.~\ref{fig:peakacc}a and \ref{fig:peakacc}b, respectively.

In Fig.~\ref{fig:peakacc}a we present four loss-of-equilibrium $a(H)$ curves calculated for different values of $C_{\rm c}>0$ and the corresponding values of $C_{\rm L}^*$, applying fixed values $X=5$ and $C_{\rm g}=0$. The graph illustrates that the peak acceleration $a_{\rm m}$ is larger for a larger value of $C_{\rm c}$, i.e., larger $C_{\rm L}^*$, and that $a_{\rm m}\rightarrow 0$ for $C_{\rm c}\rightarrow 0$.
The value of $a_{\rm m}$ increases if larger values of $X$ or lower values of $C_{\rm g}$ are applied. Note that in all cases shown in Fig.~\ref{fig:peakacc}a the eruption starts at the same height, very close to $H=1$. This is consistent with conclusions drawn by \citep[e.g.,][]{chen89,vrs90} and confirmed by observations \citep[e.g.,][]{vrs91,chen03}. This property is a consequence of the hoop-force behaviour, achieving a maximum at $H=1$, since at this height the radius of curvature attains minimum, $\tilde{R}_{\rm t}=1$.

For comparison, besides the four $a(H)$ curves based on the Kuperus-Raadu configuration ($C_{\rm c}>0$) shown in Fig.~\ref{fig:peakacc}a, one example of the Kippenhahn-Schl\"uter case is presented (gray-dotted curve; $C_{\rm c}$\,=\,$-700$\,m\,s$^{-2}$). It reveals a completely different behaviour than the Kuperus-Raadu configuration. First, the equilibrium point is shifted to much lower heights than in the Kuperus-Raadu case. Furthermore, the metastable form of the $a(H)$ curve is possible only for $C_{\rm g}$\,$>$\,0, and in that case there is always an upper-equilibrium position present (the x-axis intercept of the $a(H)$ curve located at $H$\,$\approx$\,1 in the presented case). Thus, the Kippenhahn-Schl\"uter configuration can result only in failed eruptions, and will not be considered hereafter.

To illustrate the effect of gravity, Fig.~\ref{fig:peakacc}b shows four loss-of-equilibrium $a(H)$ curves for different values of $C_{\rm g}$ and the corresponding values of $C_{\rm L}^*$, applying fixed values $d$\,=\,70\,Mm, $X$\,=\,5, and $C_{\rm c}$\,=\,1.3\,km\,s$^{-2}$. The three curves calculated with $C_{\rm g}\neq 0$ show an upper equilibrium point, unlike the curve calculated with $C_{\rm g}$\,=\,0. Thus, in $C_{\rm g}$\,$\neq$\,0 cases, the eruption should stop at an upper equilibrium. As a matter of fact, after loosing equilibrium, the rope should accelerate upwards, reach maximum speed at the upper equilibrium, after which it should be decelerated until being stopped. Then, the restoring force at the upper equilibrium would start accelerating it downwards -- in the absence of drag the rope should oscillate around the upper equilibrium position. Taking into account the drag effect, the oscillations would be damped until the rope settles at the upper equilibrium \citep[for observations see, e.g.,][]{vrs90osc,mrozek08}.

In Fig.~\ref{fig:peakacc}b we also show the case $C_{\rm g}$\,=\,0.3\,m\,s$^{-2}$ with $C_{\rm L}^*$\,=\,144\,m\,s$^{-2}$ (red-dotted curve), to compare it with the curve calculated for the same $C_{\rm L}^*$\,=\,144\,m\,s$^{-2}$ and $C_{\rm g}$\,=\,0 (red-full curve). The graph shows that the gravity effect shifts the $a(H)$ curve downwards, and since the effect is weaker at larger heights due to decreasing $g$, this leads to the formation of the upper equilibrium position. In this respect, let us also note that the gravity effect weakens with increasing $d$, i.e., it is smaller for larger ropes. For example, for $d$ two times larger than the applied value of $d$\,=\,70\,Mm, the upper equilibrium disappears in the $C_{\rm g}$\,=\,0.3\,m\,s$^{-2}$ case.

The effect of the decreasing background magnetic field is illustrated in Fig.~\ref{fig:peakacc}c, where we show $a(H)$ curves calculated using again $C_{\rm L}^*$\,=\,144\,m\,s$^{-2}$, $C_{\rm c}$\,=\,1.3\,km\,s$^{-2}$, $X$\,=\,5, and $C_{\rm g}$\,=\,0, but now applying several different values of the exponent $n$ in the dependence $B_{\rm c}$\,=\,$B_{\rm 0c}\tilde{R}_{\rm t}^{-n}$.
The red curve, representing the case $n$\,=\,0, is the same as the red curves in Figs.~\ref{fig:peakacc}a and \ref{fig:peakacc}b. The graph shows that the peak acceleration increases with increasing $n$, i.e., it is larger for eruptions occurring in coronal regions where the decrease of the magnetic field is steeper. The rate of decrease is dependent on the flux distribution at the photosphere. Furthermore, the acceleration peak systematically shifts to lower heights.

Although the effects presented here are similar to those governing the torus instability, they does not describe the same physical mechanism. If one would like to consider the torus-instability effect, the pressure-gradient term in the hoop force, i.e., the first term in Equation~(\ref{Eq:net2}), should be replaced by a new term, estimated using a new difference between the magnetic field strength at the inner and outer boundary of the torus. The difference can be estimated from the derivative of $B_{\rm c}=B_{\rm 0c}\tilde{R}_{\rm t}^{-n}$,
and from that one gets the pressure-gradient term of the form $2\,nC_{\rm L}/\Lambda\tilde{R}_{\rm t}^{n+1}$. Note that this approach is different from that usually applied \citep[e.g.,][]{kliem06}, where the equation of motion is derived employing the virtual-work principle. Consequently, the equation of motion is of a somewhat different form than that used in torus instability studies. For example, the ``standard'' torus instability equation of motion includes explicitly the self-inductance, whereas it is herein included implicitly. In addition, Equation~(\ref{Eq:net2}) includes the diamagnetic effect, Amp\'ere force, and gravity, which are not usually considered in torus instability studies.

The torus instability effect is illustrated in Fig.~\ref{fig:peakacc}d, where $a(H)$ curves for several values of $n$ are presented. Comparing this graph with the graphs presented in Fig.~\ref{fig:peakacc}a-c, one finds a considerably different behaviour of the flux rope. First, the loss-of-equilibrium height $H^*$ is increasing with decreasing $n$, and at the same time, the peak acceleration $a_{\rm m}$ decreases. Second, in the case of $n>1$ the $a(H)$ profile is more sharply peaked than profiles shown in Fig.~\ref{fig:peakacc}a-c, and the peak accelerations are higher. Thus, eruptions driven by the torus instability should be more impulsive than those driven by the hoop force that represents a form of the kink instability.
Finally, note that an eruption is not possible if the value of $n$ is too low, due to the trend $a_{\rm m}\rightarrow 0$ and $H^*\rightarrow \infty$ for decreasing $n$. For example, for the set of parameters used for Fig.~\ref{fig:peakacc}d, the critical value of $n$ is $\approx 0.8$.

The analysis presented here shows that under the conditions considered it is difficult to get accelerations higher than 1000\,m\,s$^{-2}$.
For example, even taking $C_{\rm L}$ as high as 1000 m\,s$^{-2}$ in the hoop-force option (at the same time requiring extremely large $C_{\rm c}$\,=\,100\,km\,s$^{-2}$),
one gets a peak acceleration up to 600 m\,s$^{-2}$.
This limitation is related to the conservation of magnetic flux associated with the electric current in the eruptive structure, in conjunction with the inductivity effects. As previously demonstrated, these effects cause a decrease of the current as the size of an eruption increases, which in turn leads to a rapid decrease of the driving Lorentz force. Thus, in order to get a really high acceleration, the flux-conservation constraint has to be broken. This can be attributed either to an abundant flux emergence, as proposed by \citet{chen89}, or by eruption-related reconnection resulting in the flare energy release, as suggested by \citet{vrs08angeo}. Since in the former case the emergence rate should be much higher than ever observed, the latter option seems to be more appropriate \citep{kunkel10}. It is also important to note that since the open field configuration holds more energy than the closed state,
it has been postulated that magnetic reconnection is necessary for a successful eruption to occur \citep{aly84,sturrock91}.

The role of reconnection in the dynamics of a flux rope eruption was analysed by \citet{vrs08angeo} and \citet{vrs16}, providing an insight into the physics of the CME/flare relationship. Reconnection results in two effects that can significantly enhance and prolong the CME acceleration. First, reconnection supplies the rope with additional poloidal flux \citep[see Fig. 4 in][]{vrs08angeo}, which enhances the hoop force, providing higher peak accelerations.
Furthermore, the poloidal-flux supply reduces the inductive decrease of the axial current in the later phases of the eruption, i.e., the acceleration phase is prolonged.
The second important effect of reconnection is the weakening of the magnetic tension of the arcade field overlying the flux rope by detaching this field from the photosphere.

In \citet{vrs08angeo} and \citet{vrs16} the reconnection effects were analysed assuming that reconnection starts simultaneously with the force imbalance, attains maximum rate when the rope is at a certain height, and ends at twice this height.
Various combinations of the height range and peak reconnection rate were considered to find out how these parameters affect the eruption kinematics \citep[see Fig. 2 in][]{vrs16}.
It was demonstrated that the acceleration time profile is tightly synchronized with the reconnection-rate time profile, whereas the velocity time profile is synchronized with the cumulative reconnected flux. This explains the observed synchronization of the CME acceleration phase and the impulsive phase of the associated flare \citep{kahler88,neupert01,zhang01,shan03,vrs04darije,zhang04,zhang06,maricic07,temmer08,temmer10}. It was also shown that the peak acceleration is higher for higher peak reconnection rate. Consequently the maximum speed of the eruption is higher for larger total reconnected flux, consistent with observations \citep{qiu05}. It was also demonstrated that the highest observed accelerations, reaching $\approx 10$\,km\,s$^{-2}$,
could be achieved if the total reconnected flux is a few times larger than the initial poloidal flux of the rope, corresponding to $\approx10^{22}$\,Mx. Such amounts are frequently observed in flares \citep[e.g.,][]{qiu04,qiu05,qiu07,miklenic07,miklenic09}.

The sequence of events presented here can be brought together in a relatively simple eruption scenario. If the coronal structure contains a flux rope and this rope is characterized by weak electric current and/or low degree of twist (i.e., low values of $C_{\rm L}$ and $X$, respectively), it is stable. If the rope is displaced from the equilibrium position, it will oscillate around the equilibrium position.
If $C_{\rm L}$ or $X$ (or both) start to increase due to photospheric twisting, shearing, or flux emergence, the structure gradually evolves through a series of quasi-equilibrium states. The equilibrium height in the $a(H)$ curve increases, which is reflected in a slow rise of the structure. At a given amount of twist, the structure becomes metastable, and if pushed strongly enough to reach the point of unstable equilibrium, it will erupt.

If the structure is sufficiently sheared or twisted, the gradual rise of the structure caused by increasing $C_{\rm L}$ and $X$ ends up with a loss of equilibrium, and consequently, an eruption. The threshold value of $X$ (the ratio of poloidal to axial field)
is lower for structures characterized by larger $C_{\rm L}$.
It is important to note that observations of the flux swept out by flare ribbons, compared to in situ flux measurements of the associated magnetic cloud,
indicate that the pre-eruption structure is weakly twisted, consistent with EUV and soft X-ray observations. The poloidal flux is added via the flare reconnection \citep{qiu07, hu14}.
Under certain circumstances, when the $a(H)$ curve shows the presence of an upper equilibrium, the eruption might be halted, showing damped oscillations around the upper equilibrium position.
Otherwise, the eruption continues into the heliosphere. The acceleration is stronger, and speed higher, for larger values of $C_{\rm L}$ and $X$. Since $C_{\rm L}$ is related to the current ($C_{\rm L}\propto I^2$) and $X$ with the self-inductance ($L\propto X$), this is equivalent to saying that the acceleration and speed are higher for structures containing larger amount of free energy $W=LI^2$. Furthermore, since $C_{\rm L}$ can be also expressed as $C_{\rm L}\propto B^2/\rho$, we find that acceleration and speed are higher for structures characterized by a higher Alfv\'en speed. This is consistent with the order-of-magnitude relationships anticipated at the beginning of this section.

To conclude this sub-section, the analysis presented demonstrates that events characterized by $a_{\rm max}$\,$>$\,1\,km\,s$^{-2}$ can occur only if the eruption is accompanied by strong magnetic reconnection below the erupting rope. Indeed, the most powerful/impulsive eruptions are associated with powerful/impulsive flares, supporting such a conclusion. Note that the reconnection process can go on for a long time, on the order of a day, as evidenced by growing post-eruption loop systems \citep[e.g., ][and references therein]{veronig06,vrs06} and signatures of current sheets in the wake of the CME \citep[e.g.,][]{ko03,webb03,lin05,ciaravella08,vrs09sheet}. Finally, the acceleration can be additionally enhanced by a transport of twist into the expanded summit of the rope \citep[][]{jockers78,browning83} and by writhing of the flux-rope axis that decreases the radius of curvature \citep[e.g.,][]{torok10}.

\subsubsection{Sympathetic Eruptions}
\label{ss:symp}

%------------------------------ FIGURE --------------------------------
\begin{figure*}[t]
\begin{center}
\includegraphics[width=1.0\textwidth]{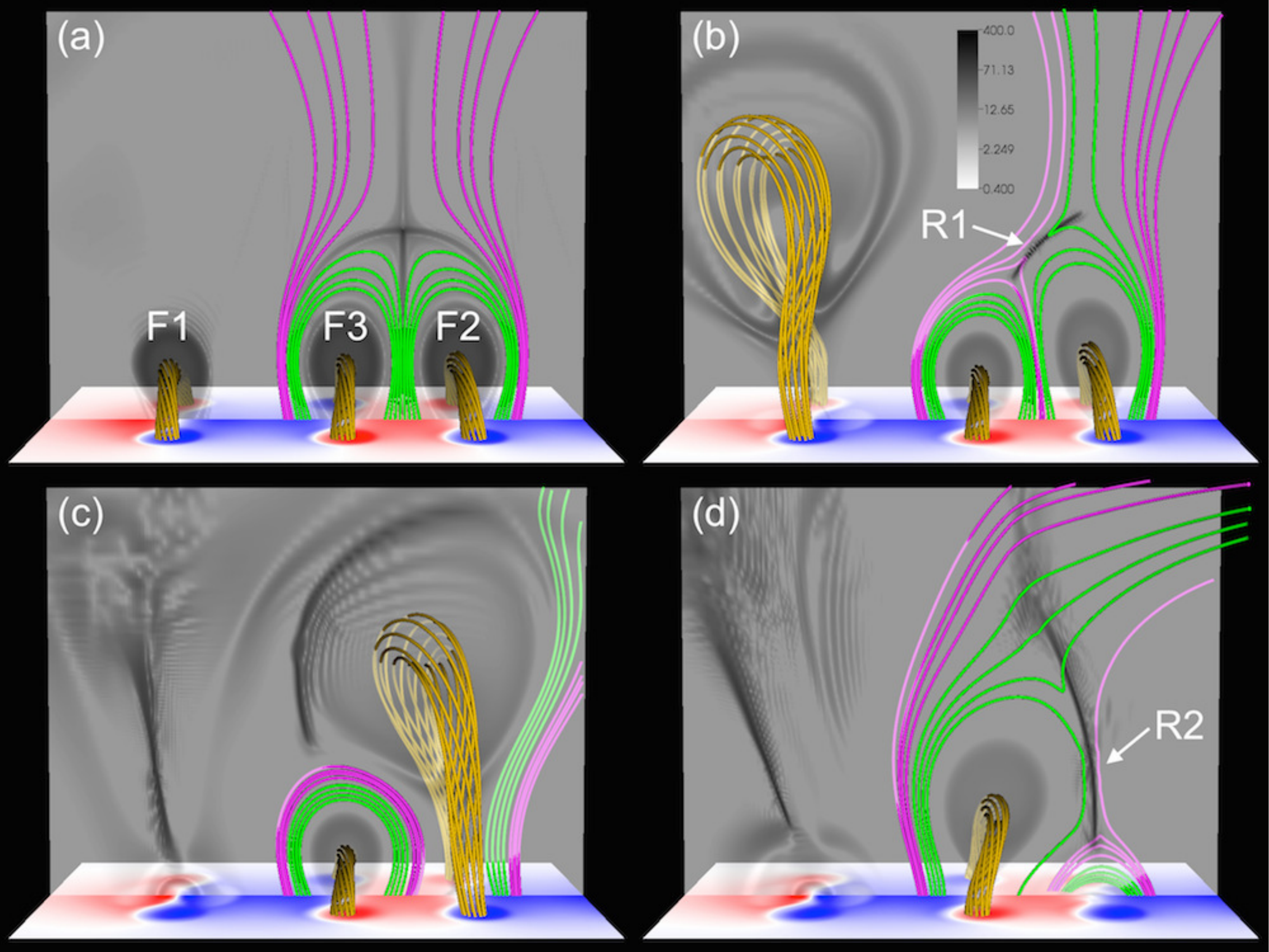}
\caption{
MHD simulation of sympathetic flux-rope eruptions by \cite{torok11}. The
eruption of F1 triggers the subsequent eruptions of F2 and F3 due to the
reconnection events R1 and R2, respectively.      
} 
\label{f:TT_symp}
\end{center}
\end{figure*}
%-----------------------------------------------------------------------

The previous discussions in this section focussed on single, isolated eruptions. However, solar eruptions can occur relatively close to each other in both space and time, and may thus interact low in the corona or later on in interplanetary space. Such interactions are discussed in Manchester et al. in this Volume; see also the recent review by \cite{lugaz16}. Here we focus on a special type of interaction, namely those in which one eruption appears to initiate another one. Such events were first mentioned more than 80 years ago \citep{richardson36}, and are now commonly referred to as {\em sympathetic} eruptions. 

Sympathetic eruptions occur within a relatively short period of time (tens of minutes to several hours) at different locations on the Sun. The individual eruptions may originate in one, typically large and complex source region \citep[e.g.,][]{liu.c09} or in different source regions that are often, but not always, adjacent.\footnote{Successive eruptions that start from essentially the same location and display very similar morphological properties are typically referred to as {\em homologous} eruptions, which we do not consider here \citep[for details on such events see, e.g.,][and references therein]{duchlev16,lugaz16}.} Research on sympathetic eruptions revolves around two main questions. Is the close timing of individual eruptions purely coincidental, or does some causal link exist between them? And if it does, what is the physical nature of such a link?  

Prior to the launch of SDO, research on the timing of flare/CME events focused mainly on the flares \citep[][]{fritzova76,pearce90,bumba93,bagala00,biesecker00,wang.h01,moon02,wheatland06}. Many of these studies considered waiting-times (i.e., the time interval between two successive events) and the location of the event, both of which are easier to establish for flares than for CMEs \citep{moon03}. These predominantly statistical studies have provided strong evidence that a causal relationship between sympathetic eruptions does indeed exist. This was supported by detailed case studies \citep[e.g.,][]{bumba93,wang.h01}. Sympathetic CMEs were studied less frequently though \citep[e.g.,][]{simnett97,moon03,cheng05,jiang.y08}, and the evidence for their existence appeared to be weaker than for flares \citep{moon03}. However, since flares and CMEs are often merely different observational manifestations of the same eruption process, this discrimination is somewhat artificial. The physical connections between sympathetic eruptions were attributed to, e.g., large-scale convective motions that led to simultaneous flux emergence in the flaring regions \citep{bumba93} or waves triggered by a previous eruption (e.g., \citealt{ramsey66,khan00}; see also \citealt{mullan76} for sympathetic stellar flares). Later it was suggested that these connections are of a magnetic nature, such as surges or jets produced by an adjacent eruption \citep{wang.h01,jiang.y08} or changes in the background field due to expansion or reconnection \citep[e.g.,][]{devore05,zuccarello09,jiang11,yang.j12}. 

The launch of SDO spurred a new series of investigations. This was triggered by the complex sympathetic eruptions that took place on 1--2 August 2010 and that were observed in unprecedented detail by AIA. \cite{schrijver11} analyzed the coronal magnetic field during the time period of the eruptions (using potential field extrapolations), and found evidence for connections between all source regions involved via structural features such as separators and quasi-separatrix layers \citep[QSLs;][]{priest92,demoulin96}. They concluded that the structural properties of the large-scale coronal field play an important role in the genesis of sympathetic eruptions \citep[see also][]{schrijver13}. Other cases of sympathetic eruptions observed by AIA can be found in \cite{shen.y12,shen.c13}, \cite{joshi.n16}, and \cite{wang.r16}. 

The 1--2 August 2010 events also inspired new modelling efforts \citep[for earlier attempts see, e.g.,][]{ding06,peng07}. Figure\,\ref{f:TT_symp} shows an MHD simulation by \cite{torok11}, which was designed to qualitatively reproduce a sub-set of these events \citep[see also][]{lynch13}. The simulation suggests two purely magnetic mechanisms for the links between successive eruptions. Figure\,\ref{f:TT_symp} panel (a) shows the initial configuration, consisting of two flux ropes (F2, F3) located in the lobes of a pseudostreamer and one rope (F1) next to it. Pseudostreamers or pseudostreamer-like configurations appear to be prone to producing sympathetic eruptions; they often harbor twin-filaments that erupt successively \citep{panasenco10}, which was the case also in the 1--2 August 2010 events. In the simulation, the eruption of F1 triggers the formation of a thin current layer around the separator line above the pseudostreamer lobes, across this current sheet breakout-like reconnection (R1) transfers flux from the right lobe to the left one (panel (b)). This reduction of flux, and thus of stabilizing magnetic tension, triggers the torus instability in F2, which subsequently erupts (panel (c)). Finally, the flare-reconnection (R2) below F2 redistributes flux from the left lobe to the growing flare arcade (panel (d)). This new flux reduction now destabilizes F3, completing the sequence of eruptions. Note that these two similar mechanisms can work independently, and their applicability is not restricted to pseudostreamers. They can work also in a quadrupolar configuration, as hypothesized by \cite{devore05}. What is required for their occurrence is merely a flux system that contains adjacent lobes overlaid by a separator line and a sufficiently strong perturbation to set things in motion. An important and yet largely unexplored question is under which conditions such reconnections will be actually strong enough to trigger a subsequent eruption \citep[see][]{torok11}. Very recently, \cite{jin.m16} undertook the first steps in answering this question, by investigating and quantifying the effects CMEs can have on adjacent and remote magnetic fields.   

The simulation shown in Fig.\,\ref{f:TT_symp} constituted a significant step forward in understanding the physical links between sympathetic eruptions. However, it was designed as a ``proof of concept'' model, using a simplified magnetic configuration. Employing instead a potential field source surface (PFSS) model based on observed magnetograms, \cite{titov12} performed a detailed topological analysis of the coronal magnetic field during the time of the 1--2 August 2010 events, and found support for the occurrence of the mechanisms suggested in \cite{torok11}. In order to further substantiate this scenario, MHD simulations with more realistic magnetic fields need to be employed to test whether sympathetic eruptions can occur also under less idealized and symmetric conditions \citep{mikic11}. Such simulations will allow us also to investigate the long-range effects of eruptions \citep[e.g.,][]{schrijver15}, as recently demonstrated by \cite{jin.m16}. Furthermore, whether other mechanisms, such as waves, can serve as a possible link between eruptions can be tested too. At present, the  observational and numerical studies that have been performed since the launch of SDO strongly support the conjecture that the mechanisms at work in sympathetic eruptions are magnetic in nature. These mechanisms, by means of reconnection, act predominantly by reducing the magnetic tension in the source region of an eruption, allowing the core field of the region to erupt.

\section{Predictability of solar eruptions}
\label{sec:predict}

Significant progress has been made in understanding the physics that underlies CME occurrence, as
well as the observational characteristics of eruptions. So, where do we stand in our ability to predict CMEs for use
in space weather forecasting?  
There are two basic goals and approaches: (1) to predict the {\em occurrence} and timing of an eruption (before it has happened) and 
(2) to predict the erupting magnetic field configuration that leaves the Sun, and its arrival time at Earth
which is related to the {\em impact} 
of an eruption. In this section, we summarise the current status of our ability to predict the 
occurrence of a CME based on observations (Section \ref{sec:pedict_obs}) and based on MHD modelling  
(Section \ref{sec:pedict_mod}).

\subsection{Predictions based on observations}
\label{sec:pedict_obs}

The evolution of the coronal and chromospheric plasma structures
in CME source regions can sometimes provide an indication of an impending eruption,
as summarised in \cite{green15}. 
However,
the timescales over which observations reveal that a CME might be imminent can vary. For example, filaments darken, 
broaden and slowly rise in the minutes prior to their
eruption \citep[for a review see][]{martin80} whereas streamers brighten over around a day 
before they lift off as a CME \citep{illinghundhausen86}. Ultimately, a variety of observations
may need to be utilised to both identify that a magnetic structure is evolving toward an eruptive configuration, 
and give a warning of when the eruption is imminent. Here we summarise the large body of observational
work that is relevant to CME prediction. 

The observational approach to predicting CMEs broadly involves two aspects; identifying the existence of a
potentially eruptive
magnetic field configuration and identifying the path to its destabilisation. Perhaps the longest-known eruptive 
magnetic field configuration is that which supports a filament. \cite{feynman1995} investigated the role
that flux emergence may play in the destabilisation of filaments and found that 19/30 of the eruptions studied 
were observed to have new flux emergence in the vicinity of the filament. The flux emerged in a 
time period of a few days before the CME. Filaments that did not have nearby flux emergence did not
erupt. The study showed that the orientation of the newly emerging flux was important, with bipoles that
were orientated in such a way as to be favourable for reconnection being key for CME occurrence. 

Sigmoids are another coronal configuration that is associated with eruptive activity. Sigmoids are 
S-shaped coronal emission structures seen in soft X-ray and EUV observations (see Section \ref{subsec:longterm_obs}), 
which form in a sub-set of CME source regions. 
These configurations are one of the most successful indicators of an eruption \citep{canfield99}
and have become synonymous with CME 
occurrence. Given the observation that sigmoidal active regions are highly likely to produce a CME, the question
can be asked ``how much time elapses between the sigmoid formation and the CME?''. 
\cite{green14} found
that around 5 to 14 hours elapsed between formation and eruption in the small sample of sigmoid regions studied.
This study used the time of the first observation of a continuous S-shaped
emission structure (rather than than an overall S shape formed in a collection of loops) 
to indicate that a sigmoid had formed.
\cite{canfield2007} found 107 active regions with overall sigmoidal shape in a study of all the
active regions imaged in partial-frame SXT data. This means that around 6\% of active regions observed in
the partial frame {\em Yohkoh}/SXT data during 1991 to 2001 were sigmoidal. Although Sterling (2000) has a
slightly more optimistic finding, stating that ``These observations indicate that the pre-eruption
sigmoid patterns are more prominent in SXRs than in EUV, and that sigmoid precursors are present in
over 50\% of CMEs.'' So, whilst sigmoids are a useful proxy, not all active regions evolve to produce a
sigmoidal configuration. In addition, sigmoids are more likely to be seen in active regions when they
are near disk centre and viewed from above, when projection effects do not mask the S shape.
As discussed in Sections \ref{subsec:longterm_obs} and \ref{subsec:shortterm_obs}, a sigmoidal emission
structure can indicate the presence of a flux rope magnetic field configuration.

The observation of a continuous sigmoid therefore gives a only few hours warning that the
region will produce a CME. Sigmoids can form in decaying bipolar active regions though,
that appear to show a systematic evolution toward this shape during their lifetime and this 
can be exploited to give a longer lead time for CME prediction. 
For sigmoids that form in decaying bipolar active regions, the process of formation takes
a couple of days. During this time the coronal structures evolve due to flux cancellation
in the photosphere. 
Flux cancellation can take the coronal field through 
three observational stages; 1) increase of shear along the polarity inversion line, 2) the formation of two sets of J-shaped loops and 
3) the formation of a sigmoid \citep{green09,green14}. Accurately identifying stage 2 would increase the 
CME forecast time by up to a day.
Sigmoids are relatively easy to identify when present in
small and bipolar regions and viewed from above. They could be harder to identify in complex 
multipolar active regions with several polarity inversion lines and strong EUV
and soft X-ray emission from adjacent structures. However, it is still possible to observe sigmoids in these environments
\citep[e.g.][]{zharkov11} where they also seem to be associated with flux cancellation.

The interpretation that some sigmoids indicate a flux rope topology has led to
studies of flux cancellation that probe the amount of flux that
builds into the rope. This can potentially be used to investigate the stability of the structure.
The flux cancellation associated with the formation of a sigmoid was studied in 
detail for the small bipolar NOAA active region 10977 \citep{green11}. The amount of flux cancelled was measured from the
time that the active region had fully emerged to the time of the first eruption from the region. 
The results showed that at the time of the eruption the ratio of flux in the rope to that in the overlying field 
of the active region had an upper value of 1:1.5.
Another study, in an active region that wasn't sigmoidal found that at the time of a filament eruption
the ratio of the flux contained in the rope to the flux of the overlying arcade field was 1:0.9 \citep{yardley16}.
However, there were previous eruptions from this region and the study wasn't able to track
the flux cancellation during the entire time during which the filament was forming, and so may represent a lower
limit to the flux that was actually contained in the rope. 
These observational results can be compared to a series of studies that use the flux rope insertion 
method \citep{vanballegooijen04} to investigate the stability of flux ropes embedded in an overlying arcade.
Stable flux ropes have been produced that have a ratio of flux in the rope to flux in the 
overlying arcade ranging from 1.15 to 1.9 \citep{bobra08,savcheva09,su09,savcheva12}. 
The \cite{savcheva12} study compared the modelled results to the observational deductions of flux rope
flux, which found that around 60\% to 100\% of the flux involved in the flux cancellation built into the rope.

Flux cancellation at the photosphere may be an indicator of the formation of an eruptive configuration
but cancellation is a ubiquitous process that takes place all over the photosphere. The challenge is to recognise
when and where this process is important for CME occurrence, which can only be done when studied in 
concert with the evolution of the coronal field.
Another aspect is that a small number of sigmoids appear more transiently in active regions.
These sigmoids appear to be formed via reconnection in the corona (rather than via flux cancellation and 
reconnection in the 
photosphere or chromosphere). Sigmoids formed in this way have been observed around 2 
hours prior to their eruption \citep{cheng14,james2017} although they tend to flash
into view and fade, only to reappear again around eruption onset making them harder to use for CME forecasting.

An alternative and complementary approach is to study the occurrence and predictability of solar flares
in order to make a CME forecast.
This approach is relevant due to the relationship
between flares and CMEs (see Section \ref{subsec:evo_eruption}) and has the added advantage that the datasets
used to study flares include observations of the photosphere, which are available over a much 
longer time period than X-ray and EUV imaging, and coronagraph observations are.
\cite{sammis2000} analysed eight years of active region observations in order to establish 
whether a relation between 
an active region's magnetic classification and the occurrence of large flares 
(for a description of the active region classes see \citealt{Hale1919}).
They find that active regions classified as $\beta \gamma \delta$ produce significantly more large flares than other 
regions of comparable size. 
Each active region of magnetic class $\beta \gamma \delta$ and a size greater than $1000$~$\mu$h revealed a probability of nearly 40\% 
of producing flares of GOES class X1 or greater. Flares of this size have around a 90\% chance of being
associated to a CME \citep{yashiro05}. 
More recently, surface magnetic parameters were combined with the amount of stored (free) magnetic energy reconstructed 
from local non-linear force-free field modelling, and tested against the flare productivity
of active regions \citep{jing2010}. Though this study revealed that the free magnetic energy clearly differentiates flaring active regions
from non-flaring active regions, the temporal variation of an region's free magnetic energy did 
not exhibit a clear pre-flare pattern.  

Studies that apply the physical knowledge of an active region's morphology and magnetic properties, 
to parameterize the solar data in order to quantify flare/CME productivity, mostly use the photospheric line-of-sight magnetic field.
However, vector field measurements can also be used. 
For the parameterizations based on the magnetic properties of active regions, one can further distinguish 
between those that relate directly to the observed photospheric field (such as magnetic flux, length and strength of main 
polarity inversion line, etc.) and those that are derived after reconstruction of the coronal field distribution 
from the vector photospheric data (such as total magnetic energy, free magnetic energy, etc.), see \cite{barnes16}. 
In general, it seems that quantities that are integrated over active region scales are stronger determinants for upcoming flare/CME
activity than are localized variations on small scales \cite[e.g. review by][]{schrijver2009}. In addition, the 
flaring history of an active region plays a decisive role \citep{falconer2012}. 
There exist statistical methods that predict flaring probabilities using solely the 
flaring history of the active region under study. This type of statistical model makes use of the information that 
the past history of flare occurrence in an active region is an important indicator of its future flare productivity 
and that the solar flare size distribution follows a power-law \citep{wheatland2004}.  
 
One of the key features used to describe the magnetic characteristics of an active region with respect to its flare/CME 
productivity is the polarity-inversion line. \cite{schrijver2007} conducted a statistical study of the magnetic properties 
of active regions associated with almost 300 M- and X-class flares, in comparison with 2,500 randomly selected active regions. 
The main finding of the study was that active regions that are the source of large flares all have a pronounced polarity 
inversion line with a strong gradient across it. Calculating the unsigned flux within 15 Mm of such strong field 
polarity inversion lines is shown to be a good indicator for the occurrence of a large flare. 

A series of papers led by Falconer used the polarity inversion line along with a photospheric proxy for the 
free magnetic energy stored in the coronal magnetic field to investigate whether there exists a relation 
to the region's CME productivity. The work used vector magnetic field data to 
derive various parameters that describe the global non-potentiality of the active region's magnetic field. 
These parameters are the length of the strongly sheared field across the 
polarity inversion line, the net electric current and the degree of twist.
They showed that all three parameters are significantly correlated 
with the active region's CME productivity \citep{falconer2002}. Note that in their study, they focussed on 
active regions with one well-defined, dominant polarity-inversion line.
In a follow up study, \cite{falconer2006} extended this work to a larger sample of active regions, and also 
extended the set of paramaters derived from the vector field data.
Three of the six parameters calculated characterize the total non-potentiality of the active region 
(length of the strong-shear segment of polarity inversion line, length of the main polarity inversion line 
with a strong field-gradient across it, and the total net electric current). Two of them describe the overall 
twist in the active region (the net-current $\alpha$ and best constant $\alpha$ derived from a linear force-free field)
using the twist parameter $\alpha$, which is determined from the net current divided by the magnetic flux.
One parameter describes the active region size (the total magnetic flux). They find that the magnetic twist and the size are not correlated, 
i.e. they are separate quantities describing the active region's characteristics, but both show a strong relation to the 
active region's CME productivity, suggesting that a combination of total flux and total twist would provide the best metric for forecasting. 
In addition, they find that the total free magnetic energy in an active region is more relevant to its CME productivity than 
is the overall twist (or its large-scale helicity) alone. 

\cite{georgoulis2007} carried out a study where only a single metric was used 
to provide a probability forecast of flaring occurrence in an active region.
The authors defined a quantity called the ``effective connected magnetic field'', $B_{\rm eff}$, 
that is calculated by connectivity matrices 
of field-line lengths and fluxes between positive and negative flux elements. 
They derived $B_{\rm eff}$ for 298 
active regions (93 X- and M-flaring, 205 non-flaring) from MDI/SOHO data over a 10~yr period during solar cycle 23, 
finding that $B_{\rm eff}$ provides a robust measure to distinguish flaring from non-flaring regions. 

\cite{leka2003a,leka2003b} conducted a thorough study into whether an active region will be 
flare-productive or flare-quiet. They derived various parameters, together with their statistical 
moments from vector magnetic field data.
These parameters include the magnetic flux distribution, horizontal spatial gradients of the field, vertical current, 
twist parameter $\alpha$, current helicity, shear angles, and photospheric excess magnetic energy. 
\cite{leka2007} examined a set of 496 active regions observed in more than 1200 magnetograms.  Statistical tests 
based on linear discriminant analysis are applied to the numerous photospheric magnetic parameters listed above, 
and compared to epochs of non-flaring. The most accurate predictor for the occurrence of large flares (defined as larger than M1), 
is found to be the total excess photospheric magnetic energy (in excess of the potential field). 
Other top performers identified are parameters 
that measure global properties of the field, i.e.\ ones that are integrated over the full active region.  
The analysis of \cite{leka2007} shows that a number of parameters are important to quantify an active region as flare-productive, 
but many of these quantities are strongly correlated. For instance, large active regions as measured by the total magnetic flux 
also tend to have large vertical currents, significant excess energy, and significant current helicity. 
Thus, the choice of which few variables to use is not unique.
The discriminant analysis in \cite{leka2007} gives a success rate of 80\% for the occurrence of flares of GOES class C1 or larger. 
This is compared to a success rate of 70\% when all regions are uniformly classified as ``flare quiet''. For flares of class M1 and larger, 
the success rate obtained is 93\%, which is only slightly larger than the rate of 91\% that is obtained for the uniform flare-quiet labelling. 
This is an intrinsic problem related to the low occurrence rate of large flares, as low event occurrences typically lead to large false alarm rates. 
Based on these outcomes, the authors conclude that 
the state of the photospheric magnetic field at any given time does not have a large influence on whether an active 
region will be flare productive or not. 
This is in line with the conclusion in \cite{schrijver2009}, that so far none of these various measures (either single or combined) 
has been identified as being particularly well correlated with the flare/CME activity of an active region. 

In order to directly compare the relative success of existing flare prediction algorithms, an interagency ``All-Clear workshop"
was held in Colorado in 2009. The outcomes of the workshop are summarised in \cite{barnes16}.
The workshop brought together people working on a number of existing algorithms. Previously, it had been difficult 
to compare the outcome of 
different flare forecasting algorithms, as they use different analysis techniques and data sets. The analysis carried out
for the All-Clear workshop was based on a common data set, using line-of-sight magnetic field and continuum intensity 
maps from MDI/SOHO, 
and applied standard verification metrics to evaluate the performance of the different methods. 
In total, 11 different algorithms were applied including the ones that are described above, 
i.e. the methods of \cite{georgoulis2007}, \cite{falconer2002,falconer2008}, \cite{leka2003a,leka2003b,leka2007}, \cite{schrijver2007} and \cite{wheatland2004}.
In addition, a number of further algorithms were included, which are now described. The method of \cite{colak2008,colak2009} is based on 
a feature-recognition technique to automatically derive the McIntosh classification from white-light images and subsequently 
uses machine learning to make a probability forecast based on the sunspot class. The method by \cite{bloomfield2012} 
uses historical flaring rates related to the McIntosh class of the source active region to make forecasts using Poisson probabilities.  
\cite{yuan2010,yuan2011} apply machine learning techniques to provide flare forecasts from characteristic magnetic parameters, 
such as the total unsigned magnetic flux, length of the strong-gradient polarity inversion line, and total magnetic energy dissipation. 
The method by \cite{mcateer2010} is based on the fractal dimension of the magnetic flux concentrations in an active region 
to determine its flare productivity. The algorithm by \cite{higgins2011} and \cite{qahwaji2008} uses automatic characterization 
of various magnetic parameters of the active region to apply flare forecasts based on cascade correlation network techniques. 
The main outcome of the comprehensive workshop study detailed in \cite{barnes16} is that no method was clearly outperforming the others. 
This may be due to the fact that there are strong correlations between the different parameters that are used in the different 
prediction schemes. For the prediction of flares of class M and larger, the set of methods studied tends to give a slightly 
positive skill score, but none of the methods achieved large skill score values. All skill scores were
$\lesssim$ 0.2. For context, a perfect forecast would have a skill score of 1 and forecasts that have positive skill score 
values indicate that that method is more successful than the reference method being used.

An interesting case study against which to examine and test ideas around predicting flares/CMEs based on 
observations is
NOAA active region 12192. This region appeared on the solar disk in October 2014 and it hosted the
largest sunspot seen by that point in solar cycle 24. From its size and complexity ($\beta \gamma \delta$) 
it was expected to be the source of a significant
amount of free magnetic energy and high-intensity flares, therefore catching the attention of space weather forecasters. 
Indeed the region produced six GOES X-class flares during its disk passage but no successful eruptions occurred
\citep{thalmann2015,sun2015}. 
Although the magnetic field of NOAA region 12192 was complex there was a 
separation of the main positive and negative magnetic polarities with time. No convergence or collision of polarities 
was seen during its first disk passage.
Small bipoles between the main spots (called ``serpentine field'') were observed indicating that the flux
emergence process was still taking place. Overall, the active region configuration was that of one main bipole, 
with at least 2 new bipolar emergences on the periphery of the sunspot region or near the centre, later in time.
The region was studied by \cite{sun2015} who compared NOAA active region 12192 to two other flare-productive regions
(11429 and 11158). They constructed NLFFF extrapolations and calculated the free energy in the coronal field,
the decay index of the coronal field and the squashing factor. They find that the region's 
core field had a low level of non-potentiality and that the overlying field was relatively strong.
The lack of colliding polarities (and therefore opportunities for magnetic reconnection) may be the reason
that a strongly non-potential field didn't build
up along any polarity inversion line in the active region 
and the lack of flux cancellation would have resulted in no tether cutting (which removes overlying restraining field).

The suggestion that CMEs may act as a valve for magnetic helicity that has accumulated in the corona
appears to lead intuitively to the idea that CMEs might occur once a
threshold in this quantity has been reached. Can this be used to forecast the occurrence of a CME?
Although this idea is by no means accepted or proven,
some work has indicated that magnetic helicity is a necessary but not sufficient condition for 
an eruption \citep{amari03b} whilst others show that magnetic helicity is not necessary for CMEs
\citep{zuccarello2009}. 
It is worth noting that CMEs originate from a wide range of coronal magnetic structures, from 
bright points with a
flux ($\phi$) of around $10^{20}$ Mx, to large active regions, where $\phi=10^{22}$ Mx, giving a potentially 
large range in magnetic helicity values (since helicity scales with $\phi^{2}$). Therefore,
there will not be a one-size-fits-all value for magnetic helicity in order to create an 
eruptive configuration. Instead,
the helicity of an eruptive configuration will vary from case-to-case. 
A study by \cite{pariat2017} takes a different approach into the potential utility of magnetic helicity
to forecast and eruption. \cite{pariat2017} study the 
evolution of stable and eruptive magnetic configurations using the flux emergence simulations of \cite{leake2013,leake2014}.
The relative magnetic helicity of the computed field is decomposed into that of the current carrying component
and that of the potential component. 
They find that the eruptive configurations are not necessarily those with the highest magnetic 
helicity value but that an eruptivity indicator exists in the ratio of these two components of the
relative magnetic helicity. Opening, for the first time, an avenue for investigating how magnetic helicity 
may be used for CME forecasting.

\subsection{Predictions Based on MHD Models}
\label{sec:pedict_mod}

MHD models provide a potentially powerful future tool in the quest for a predictive capability of CMEs.
However, at present, the MHD numerical models described in Section~\ref{sec:origin_theo_sim} cannot be 
used for predicting the time of an eruption. 
This is in part because most of these models do not provide quantitative onset criteria that could then be applied to observations or magnetic-field extrapolations of potential CME source regions. Even those that do (i.e. the kink and torus instability models) can
at present only provide rough estimates of the onset criteria. This is because
the exact thresholds of these instabilities depends on the detailed structure of the magnetic-field configuration, which is not sufficiently well known. Moreover, exact thresholds have so far only been derived for rather idealized model configurations
rather than those exhibited by the real Sun. 
A complementary approach is to create a simulation that is driven by line-of-sight magnetograms.
Using real data provides an opportunity to create a realistic simulation so that the coronal field
can be monitored for its eruptive potential. This is the approach used by \cite{gibb14} in a study of 
NOAA active region 10977 that employs
a magnetofrictional method driven by line-of-sight magnetograms. The study found
that a realistic magnetic field line reconstruction was not possible around the time that the active region
erupted suggesting that the approach captured well the destabilisation of the coronal field. Since the top boundary of the simulation box is closed, the flux rope that had formed and started to erupt wasn't able to escape the computational volume. That is, the computed field was trying to relax by ejecting the flux rope but the boundary conditions did not allow this to happen. Consequently, this led to the formation of highly twisted structures in the computational volume.

Apart from the eruption onset, one would also like to forecast the {\em arrival time} and {\em impact} at the Earth of the CME, once the 
eruption is on its way. CMEs typical take from 0.7 to 4 days to travel from the Sun to the Earth.
This is an amount of time that allows an MHD model, based on solar data, to be run.
In contrast, predicting the effects of energetic-particle events created by CME-induced shocks is much more challenging, as these particles can reach the Earth in less than an hour. We therefore consider here only to what extent MHD models can be employed for predicting the time and impact of space-weather disturbances caused by CMEs. The most important quantities for the latter are the speed, duration, plasma density, and magnetic field strength and orientation of the CME. Sophisticated models should also include the effects of the shock and the sheath that precede the CME flux-rope in fast events.  

As discussed in \cite{jin.m17}, present CME forecasting models can be divided into different categories \citep[see also][]{siscoe06,messerotti09}. The first categories contain empirical forecasting models \citep[e.g.,][]{gopalswamy01,schwenn05}, kinematic models such as the cone model \citep[e.g.,][]{zhao02} and analytical models such as the drag-model \citep{vrs13dbm}.
These models have been widely used to predict the arrival time, orientation, and speed of CMEs (and of associated shocks), but they cannot provide predictions of the plasma density and the magnetic field (though see, for example,  \citealt{mostl14} for attempts to predict the maximum magnetic field strength in an CME from its predicted speed). To this end, the use of models of the third category, namely of MHD simulations, appears inevitable.

However, MHD simulations that employ observed data and continuously model an eruption from its onset in the low corona out to 1 AU (see Manchester et al. in this Volume) are limited to using data from the Earth’s point-of-view. Therefore, the Earth-directed
component of the CME velocity is not accurately determined.
Moreover, these simulations cannot yet be performed in a constantly evolving and accurate solar wind. For this, in situ solar
wind measurements are needed that provide real data to adjust the models to.
This limits the capability of these MHD models, but efforts are being made to develop such models for future operational use \citep[e.g.,][]{jin.m17}. Presently, ENLIL \citep{odstrcil03} is the only MHD model that is used for operational forecasting (at NOAA and the UK's Met Office Space Weather Operations Centre for example). ENLIL produces estimates of the solar wind speed, density, and temperature and magnetic field, and provides 1-4 day advance warning of disturbances that will produce geomagnetic storms, such as CMEs. To be applicable in practice, simplifications had to be made to the model. For example, ENLIL simulates the solar wind plasma and magnetic field only beyond $30\,R_\odot$, i.e., the corona is neglected. To feed the model, the solar wind plasma flow and radial magnetic field at $30\,R_\odot$ are provided by the semi-empirical WSA model \citep{wang.y.m90,arge00}. If an eruption is to be simulated, the cone model is used to set up the location, direction, and speed of the CME \citep[e.g.,][]{odstrcil05}. Also, while the solar wind magnetic field is included in the model, CMEs are initiated without an internal magnetic field. Therefore, the model cannot yet be used to predict the southward component of the magnetic field during CMEs. However, observations of the CME source region can be used to give first order approximation of the 
magnetic configuration of the erupting structure. For example, through the detection of a flux rope and determination of 
the orientation of its axis and its chiral properties \citep[e.g.][]{palmerio2017}.

\subsection{Bringing observational and theoretical predictions closer together}
\label{sec:pedict_obs_mod_together}

There is much to be learnt from the complementary approaches of observational CME studies and theoretical
work, including numerical simulations. 
In terms of the most promising approaches to predicting CME occurrence ahead of time, 
combining current observational and 
theoretical knowledge, there appear to be three main avenues:

\begin{enumerate}[(i)]
\item Flux imbalance between a flux rope and overlying arcade
\item Torus instability due to the rate of decay of the arcade overlying a flux rope
\item Helicity proxy derived from the current carrying and potential field components
\end{enumerate}

For (i), observations of flux cancellation can be used to investigate the flux built into the 
rope versus that in the overlying field. However, more work is needed to understand how the shear and 
spatial extent of the cancellation in the active region influence what proportion of cancelled 
flux is built into the rope, and what the flux imbalance criteria is.
For (ii), the gradient of the overlying field can be investigated using a potential field
extrapolation, but simulations so far describe rather idealised configurations and the exact
stability criteria varies on a case-to-case basis.
For (iii), the theoretical knowledge is still developing, but early indications suggest that there might be
a critical value for the ratio of helicity in the current carrying field component to the helicity in 
potential component.
Still, none of the above are able to give the desired several days notice of the occurrence of a CME.
In addition, approaches (i) and (ii) are relevant for the flux rope concept of CMEs rather than
a sheared arcade.

\section{Summary}
\label{sec:summary}

This review aims to provide a summary of the wide range of observational and theoretical work 
on coronal mass ejections (CMEs). These eruptions are known to be a driver of space
weather, including the most severe space weather events. In response to this, 
an increasing number of developed and developing nations
provide space weather forecasts to help mitigate the risk of economic and societal disruption
resulting from the technology affected by space weather. The severe impact that results from the 
arrival of a CME of high velocity,
southward directed magnetic field and enhanced plasma density means that there is a significant
interest in forecasting these eruptions ahead of time.
 
Currently the physical processes at play after CME onset can be brought together in the standard
CHSKP model. 
This captures and explains CME observations well, once the eruption is underway. 
However, the energy required to power a CME (and any associated flare) cannot be supplied
to the corona on the timescale of the dynamic event and instead it must be built up and
stored in the coronal magnetic field in the hours or days beforehand. 
We need to extend the standard model further back in time to understand the physical
processes that create an eruptive magnetic field configuration and the associated currents, 
as well as the processes that trigger and drive the eruption.
CME occurrence requires more than just estimating the free magnetic energy stored in the coronal field - 
the free energy is a necessary but not sufficient condition. The exact magnetic field configuration 
and its stability need to be better characterised. 

Many flare/CME forecasts currently focus on using a snapshot of the configuration of the photosheric magnetic field to provide 
a probabilistic forecast. Improvements to this technique are limited by the physical origins of a 
CME being related to the stability of the coronal field, when a trigger switches a driver on, 
which is unlikely to be captured in the photospheric data.
CME drivers themselves can be monitored over many days using both photospheric and coronal
data. But this must be supplemented with data to reveal the likelihood of a CME occurring, using 
either a NLFFF extrapolation or
an eruptivity signature(s) determined using atmospheric observations.
With no regular direct measurements of the coronal
magnetic field, observations are limited to inferences obtained from plasma structures.

To move on from probabilistic forecasting, to the capability of being able to predict CME onset time, 
we first need a clear understanding of the criteria that encapsulates
the eruptive status of the magnetic field. At what point does a CME driver switch on?
Parametric modelling studies that can identify the loss of stability criteria 
are required and this could be determined from more realistic magnetic field configurations
and their evolution.
Developing the capability to conduct data driven simulations, relevant for active regions 
from their birth to their dispersal and decay into the background, field will greatly help this aim.

\begin{acknowledgements}
LMG acknowledges support through a Royal Society University Research Fellowship and through the Leverhulme Trust Research Project Grant 2014-051.
T.T. acknowledges support by NASA's LWS and H-SR programs and by NSF grants AGS-1249270, AGS-1348577, and AGS-1135432 (Sun-2-Ice). BV acknowledges the support by the Croatian Science Foundation under the project 6212 ``Solar and Stellar Variability''. AV acknowledges support by the Austrian Science Fund (FWF): P27292-N20.
We acknowledge the International Space Science Institute (ISSI) in Bern Switzerland for their
generous support and for travel and accommodation.
\end{acknowledgements}

%=====================================================================
% Modified by TT.
% I added the three style files, which were missing, and created a bib
% file 'references.bib'.
%
% BibTeX users please use one of
\bibliographystyle{spbasic}      % basic style, author-year citations
\bibliography{references}   % name your BibTeX data base

%% Non-BibTeX users please use
%\begin{thebibliography}{}
%%
%% and use \bibitem to create references. Consult the Instructions
%% for authors for reference list style.
%%
%\bibitem{RefJ}
%% Format for Journal Reference
%Author, Article title, Journal, Volume, page numbers (year)
%% Format for books
%\bibitem{RefB}
%Author, Book title, page numbers. Publisher, place (year)
%% etc
%\end{thebibliography}
%=====================================================================

\end{document}